\documentclass[sigplan]{acmart}

\AtBeginDocument{%
  }

\renewcommand\footnotetextcopyrightpermission[1]{}

\setcopyright{acmlicensed}
\copyrightyear{2018}
\acmYear{2018}
\acmDOI{XXXXXXX.XXXXXXX}

\acmConference[Conference acronym 'XX]{Make sure to enter the correct
  conference title from your rights confirmation emai}{June 03--05,
  2018}{Woodstock, NY}
\acmISBN{978-1-4503-XXXX-X/18/06}

\begin{document}

%\title{TPipe: Temporal-Locality-Aware Pipeline Parallelism for Memory-Efficient LLM Training}

\title{Enhancing Memory Efficiency in Large Language Model Training Through Chronos-aware Pipeline Parallelism}


\author{Xinyuan Lin}
\authornote{Both authors contributed equally to this research.}
\email{linxinyu22@mails.tsinghua.edu.cn}
%\email{webmaster@marysville-ohio.com}
\affiliation{%
  \institution{Dept. of Electronic Engineering \\ Tsinghua University  
  }
  \city{Beijing}
  \country{China}
}

\author{Chenlu Li}
\authornotemark[1]
\email{cll@birentech.com}
\affiliation{%
  \institution{BirenTech}
  \city{Shanghai}
  \country{China}
}

\author{Zongle Huang}
\email{huangzl23@mails.tsinghua.edu.cn}
\affiliation{%
  \institution{Dept. of Electronic Engineering \\ Tsinghua University  
  }
  \city{Beijing}
  \country{China}
}

\author{Chunyu Wang}
\email{cywang@birentech.com}
\affiliation{%
  \institution{BirenTech}
  \city{Shanghai}
  \country{China}
}

\author{Bo Xiao}
\email{yxun@birentech.com}
\affiliation{%
  \institution{BirenTech}
  \city{Shanghai}
  \country{China}
}

\author{Huazhong Yang}
\email{yanghz@tsinghua.edu.cn}
\affiliation{%
  \institution{Dept. of Electronic Engineering \\ Tsinghua University  
  }
  \city{Beijing}
  \country{China}
}

\author{Shishi Duan}
\email{burnessduan@birentech.com}
\affiliation{%
  \institution{BirenTech}
  \city{Shanghai}
  \country{China}
}

\author{Yongpan Liu}
\email{ypliu@tsinghua.edu.cn}
\affiliation{%
  \institution{Dept. of Electronic Engineering \\ Tsinghua University  
  }
  \city{Beijing}
  \country{China}
}

% %%
% %% By default, the full list of authors will be used in the page
% %% headers. Often, this list is too long, and will overlap
% %% other information printed in the page headers. This command allows
% %% the author to define a more concise list
% %% of authors' names for this purpose.

\begin{abstract}
Scaling LLMs through increased parameters and sequence lengths imposes significant storage pressure on LLM training. While expensive HBM requires advanced packaging for capacity scaling, existing pipeline parallelism strategies prioritize computational bubble reduction over memory efficiency. Consequently, existing methods fail to address the increasing storage requirements for ever-growing LLMs.

This work presents TPipe, a temporal-locality-aware pipeline parallelism for memory-efficient LLM training. The core insight of TPipe is to treat HBM as a fast but small 'cache,' optimizing and exploiting temporal locality within LLM. TPipe solution comprises
three components: T-Pipe, T-Recomp, and T-Offload. T-Pipe, a pipeline scheduling strategy, is used to reduce the extrinsic overhead that disrupts the temporal locality of activations. T-Recomp and T-Offload are utilized to efficiently harness the intrinsic temporal locality of the activations and weights in DNN. Experiment results show that TPipe can expand the trainable model size by 2.4x while maintaining comparable throughput, achieving 1.5x better than the 1F1B strategy combined with recomputation.
\end{abstract}


%  Xinyuan Lin add
\iffalse



\fi


\received{20 February 2007}
\received[revised]{12 March 2009}
\received[accepted]{5 June 2009}

\maketitle

\section{INTRODUCTION}

% In recent years, Large Language Models (LLMs) have demonstrated exceptional performance across diverse domains~\cite{LLM}, which is largely attributed to the substantial memory requirements during pre-training. It is governed by two primary factors: \textbf{Parameter Scaling}: The scaling law~\cite{ScalingLaw} posits that LLMs with increased parameter counts yield superior performance. This trend is exemplified by the rapid expansion from GPT-1 (0.12B parameters)~\cite{GPT1} to Llama 3.1 (405B parameters)~\cite{LLAMA3}, representing a more than 3000-fold increase in model size within a mere six-year span. \textbf{Sequence Length Extension}: Longer sequence lengths have proven instrumental in enhancing model capabilities, particularly in multimodal applications~\cite{Multimodal} that necessitate the processing of extensive video and audio sequences. Consequently, the memory requirements for activations, weights, gradients, and optimizer states keep soaring.

In recent years, Large Language Models (LLMs) have demonstrated exceptional performance across diverse domains~\cite{LLM}, leading to an urgent demand for storage capacity during pre-training. This demand can be attributed to two main reasons outlined below: \textbf{Parameter Scaling}: The scaling law~\cite{ScalingLaw} posits that LLMs with increased parameter counts yield superior performance. This trend is exemplified by the rapid expansion from GPT-1 (0.12B parameters)~\cite{GPT1} to Llama 3.1 (405B parameters)~\cite{LLAMA3}, representing a more than 3000-fold increase in model size within a mere six-year span. \textbf{Sequence Length Extension}: Longer sequence lengths have proven instrumental in enhancing model capabilities, particularly in multimodal applications~\cite{Multimodal} that necessitate the processing of extensive video and audio sequences. Consequently, the memory requirements for activations, weights, gradients, and optimizer states keep soaring.

% Contemporary GPU training systems utilize high-bandwidth memory (HBM) for runtime variable storage, but HBM capacity expansion faces significant challenges. First, HBM integration requires advanced packaging techniques such as through-silicon via (TSV)~\cite{TSV} and microbump~\cite{microbump}, resulting in substantial manufacturing costs. For instance, HBM accounts for 50-60\% of the total costs in NVIDIA's H100 SXM5 module~\cite{H100}. Moreover, physical constraints pose severe restrictions. Increasing the count of stacking DRAM die within the restricted packaging height (720 um) necessitates hybrid bonding~\cite{HybridBonding}, whose yield optimization are extremely challenging~\cite{HybridBonding_Problem}.

Contemporary GPU training systems utilize high-bandwidth memory (HBM) for runtime variable storage, but it now faces significant challenges. First, HBM integration requires advanced packaging techniques such as through-silicon via (TSV)~\cite{TSV} and microbump~\cite{microbump}, resulting in substantial manufacturing costs. For instance, HBM accounts for 50-60\% of the total costs in NVIDIA's H100 SXM5 module~\cite{H100}. Moreover, increasing the count of stacking DRAM die within the restricted packaging height (720 um) necessitates hybrid bonding~\cite{HybridBonding}, whose yield optimization~\cite{HybridBonding_Problem} is extremely challenging.

\begin{figure}[htbp]
    \centerline{
    \includegraphics[width=0.48\textwidth]{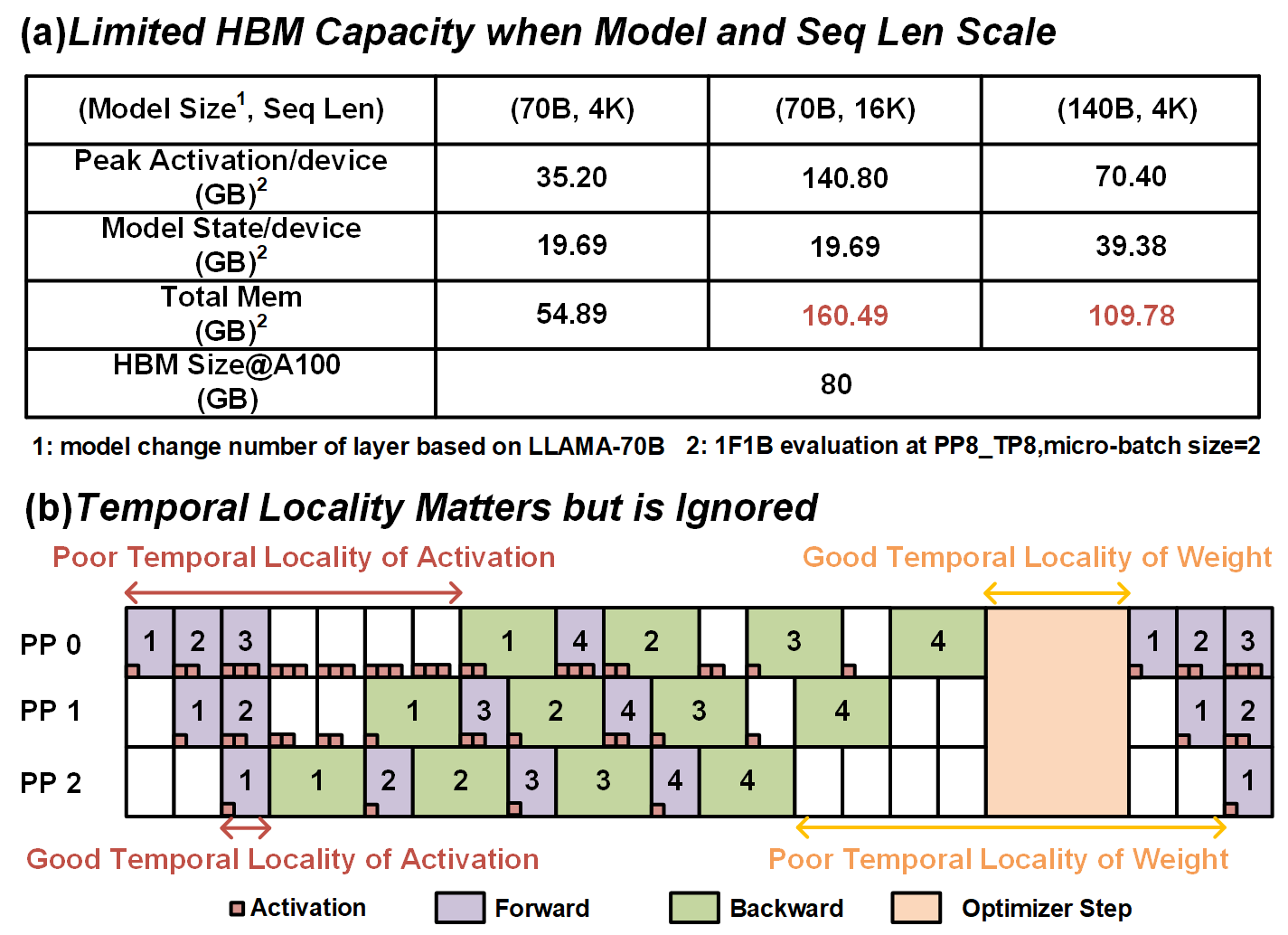}
    }
    \caption{(a) limited HBM capacity is difficult to meet the demands of next-generation LLM. (b) existing pipeline parallelism (such as 1F1B) ignores temporal locality.}
    \label{fig:1}
\end{figure}

Therefore, LLM training necessitates memory-efficient scheduling strategies while existing Pipeline Parallelism (PP) predominantly emphasizes performance optimization and overlooks memory constraints. Currently, most works focus on pipeline bubble reduction through fine-grained task division~\cite{GPipe,DAPPLE,Interleave-1F1B,Hanayo,ZeroBubble} or utilizing tasks in other dimensions~\cite{Chimera,Mixpipe}. However, these methods fail to address activation storage imbalances across pipeline stages (as shown in Fig. ~\ref{fig:1}(b)) and may inadvertently increase activation and weight storage requirements. Few works notice this issue and propose offloading activations to CPUs~\cite{MegatronKwai} or additional GPUs~\cite{Bpipe}. However, unlike the offloading of optimizer states, which only happens once per mini-batch, activation offloading is more frequently conducted across different micro-batches, thus raising a higher demand for offloading bandwidth.

% Moreover, current LLM training predominantly employs hybrid parallelism, yet existing PP strategies lack optimal compatibility with other memory-efficient scheduling approaches. The ZeRO series~\cite{ZeRO} represents state-of-the-art data parallelism (DP) techniques, distributing model states across DP machines. In hybrid DP and PP configurations, distributing optimizer states (ZeRO-1) necessitates inter-DP rank communication only at \textbf{mini-batch} granularity. However, further distributing weight gradients (ZeRO-2) and weights (ZeRO-3) requires more frequent communication at \textbf{micro-batch} granularity, imposing higher bandwidth demands between DP ranks. Consequently, hybrid DP-PP implementations~\cite{DeepSpeed} typically utilize only ZeRO-1, leaving substantial potential for reducing weight and gradient storage. When designing PP strategies that are more compatible with ZeRO-2/3, such as Breadth First PP~\cite{BreadthFirstPP} and ZeROPP~\cite{ZeroPP}, these approaches reduce DP bandwidth requirements at the cost of a significant increase in activation memory, leaving no perfect solution.

Existing PP strategies lack memory efficiency due to their failure to address an intrinsic property of Deep Neural Network (DNN): \textbf{temporal locality}. In these networks, forward computation precedes backpropagation, causing activations in shallower layers to be generated early but released late, thus incurring a worse temporal locality. In pipeline parallelism, where layers are distributed across machines, the first stage that contains the shallowest layer experiences peak activation storage.

Unlike previous work, we notice the issues mentioned above and introduce the concept of temporal locality to pipeline parallelism for the first time. We propose TPipe, which enables the training of models 2.4 times larger than 1F1B~\cite{DAPPLE} on the same hardware with comparable throughput while maintaining compatibility with various memory-efficient scheduling strategies. The TPipe solution comprises three components: T-Pipe, T-Recomp, and T-Offload.

\textbf{T-Pipe} minimizes peak activation storage by reducing micro-batch execution time, thus expediting the backward pass that consumes activations. This, in effect, optimizes activation temporal locality. \textbf{T-Recomp} exploits the poorer temporal locality of shallower layer activations and discards them from HBM by selectively recomputing to achieve higher efficiency. \textbf{T-Offload} takes advantage of poorer temporal locality of deeper layer weight by offloading optimizer update to CPU during bubbles created in T-Pipe. Since this process overlaps with warmup/cooldown phase, it lowers demand on offload bandwidth and CPUs. \textbf{These strategies co-design based on temporal locality variations in pipeline parallelism, which remain underexplored in existing methods.}

Our main contributions are as follows.
\begin{itemize}
    \item We found that temporal locality in DNN is the primary cause of imbalanced activation storage in PP, constraining trainable model sizes on current devices.
    % \item For activation memory savings, we introduce temporal locality into PP schedule (T-Pipe), and recomputation (T-Recomp).T-Pipe eliminates unnecessary intervals optimizing the temporal locality of activation, reducing peak activation storage to 75\% of 1F1B while maintaining similar MFU (Section X.X). T-Recomp focuses on selectively recomputing activation in shallower layers, outperform 1.5x memory efficiency than current recomputation. 
    % \item For Model State storage capacity, we show that a ZeRo-2-compatible PP could built based on T-Pipe (Section X.X). Additionally, we have design the Offload strategy based on temporal locality (T-Offload), which can offload optimizer update of deeper layers to CPU (Section X).
    \item For activation memory savings, we introduce temporal locality into the PP schedule (T-Pipe) and recomputation (T-Recomp). T-Pipe eliminates unnecessary intervals, optimizing the temporal locality of activation (Section \ref{T-Pipe}). T-Recomp discards activation with poor temporal locality from HBM by selectively recomputing to achieve higher efficiency (Section \ref{T-Recomp}).
    \item For model state storage, we designed the offload strategy based on temporal locality (T-Offload), which can discard model states with poor temporal locality from HBM by offloading optimizer updates of deeper layers to CPU during bubbles created in T-Pipe (Section \ref{T-Offload}).
    \item End-to-end evaluation on TPipe is carried out on a cluster comprising up to 64 GPUs. Experiments show that T-Pipe can expand the trainable model size by 2.4x while maintaining comparable throughput, achieving 1.5x better than the 1F1B strategy combined with recomputation.
\end{itemize}

\section{BACKGROUND AND MOTIVATION}
In this section, we will introduce pipeline parallelism (PP) (Section \ref{PP}), typical recomputation strategies and offloading strategies (Section \ref{Recomp}). We will then discuss the motivation for leveraging temporal locality (Section \ref{Motivation}) and review how prior work has addressed memory imbalance challenges in PP (Section \ref{Motivation2}).

To facilitate clearer explanations throughout the paper, we will use the following symbols.

$\bullet$ $m_a$:memory consumption for activation in whole Neural Network(exclude input embedding layer and language modeling head) \\
\indent $\bullet$ $P$:\# of pipeline stages\\
\indent $\bullet$ $m$:\# of micro-batches within a training iteration \\
\indent $\bullet$ $T_{fwd}/T_{bwd}$: execution time of forward/backward pass in one micro-batch

\subsection{Pipeline parallelism}\label{PP}

\begin{figure}[htbp]
    \centerline{
    \includegraphics[width=0.48\textwidth]{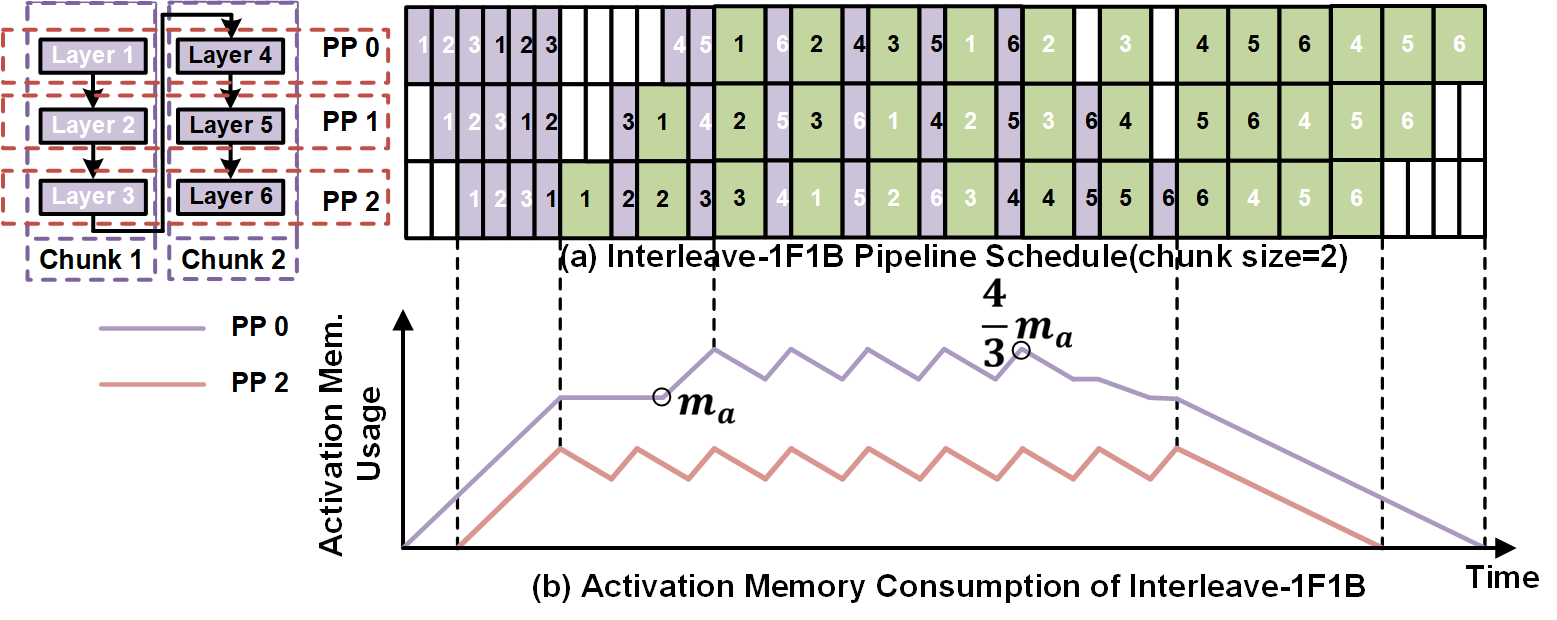}
    }
    \caption{(a) Interleave-1F1B (b) and its memory usage}
    \label{fig:2}
\end{figure}

% Pipeline Parallelism (PP) divides the model into multi-layer blocks mapped across different PP stages. These stages work together to train all samples within a mini-batch, enabling mini-batch Stochastic Gradient Descent. During the forward pass, samples in each mini-batch begin computation at the first stage, with subsequent blocks executing their computations in a pipelined manner until the final forward computation completes at the last stage. In the backward pass, gradients for each mini-batch flow in the reverse direction. As a result, PP only requires a single tensor to be sent and received per block, significantly reducing communication bandwidth demands. This approach is often deployed across nodes and is a key component in hybrid parallelism.

Pipeline Parallelism (PP) divides the model into multi-layer blocks mapped across different PP stages. These stages work together to train all samples within a mini-batch, enabling mini-batch Stochastic Gradient Descent. Forward computations in each mini-batch propagate sequentially through stages until completion, while gradients flow inversely during backward passes. As a result, only one layer of intermediate activation/gradient needs to be sent and received per block, significantly reducing communication bandwidth demands.

Mini-batch-level PP suffers from significant pipeline bubbles, commonly addressed by splitting mini-batches into micro-batches. Dividing workloads lets the first stage proceed from initial to subsequent micro-batches without idling, thus reducing pipeline bubbles~\cite{GPipe}. When the number of micro-batches increases, the classic 1F1B strategy~\cite{DAPPLE} is often used, as shown in Fig. ~\ref{fig:1}(b). During the steady phase, each forward computation block follows a backward block completion, preventing concurrent forward passes during warm-up that elevate peak memory usage. However, this approach creates an imbalance in activation memory usage across PP stages. In a single micro-batch, the first stage initiates forward computation first but finalizes backward passes and release memory last. As a result, the first stage reaches peak activation memory of $m_a$, while the last stage’s activation memory usage is only $\frac{m_a}{p}$. Fig. ~\ref{fig:1}(b) illustrates that tasks with dependencies between adjacent PP stages execute sequentially, leading to non-overlapping point-to-point (P2P) communication. Some methods mitigate this by inserting independent tasks from other micro-batches between dependent blocks to enable communication overlap. This delay caused by dependency is called the "interval" in this paper.

% Blocks can be broken down into smaller tasks to reduce pipeline bubbles further, though this may increase memory requirements. The Interleaved-1F1B approach~\cite{Interleave-1F1B} divides blocks into more chunks and schedules multiple chunks launches during the warmup phase, as shown in Fig.~\ref{fig:2}(a). When the workload is divided into v chunks  (assigning layer 1, $p+1$, …, $ (v - 1)p+1$  to stage 0; layer 2, $p+2$, …, $(v-1)p+2$ to stage 1, and so forth), the bubble size is reduced to $\frac{1}{v}$ of that in the traditional 1F1B, but peak activation memory increases to $m_a(1+\frac{p-1}{pv})$. The Zero Bubble approach~\cite{ZeroBubble} goes a step further by splitting the backward pass (BP) into two phases: activation gradient computation (BPA) and weight gradient computation (BPW). This arrangement theoretically enables a bubble-free pipeline. However, the activation gradients generated by BPA are required to be retained until BPW is completed, which raises memory demands. Additionally, splitting the BP disrupts the original reuse of activation gradients for all-gather operations between BPA and BPW, and reduces the overlap between computation and communication.

Blocks can be broken down into smaller tasks to reduce pipeline bubbles further, though this may increase memory requirements. Interleaved-1F1B ~\cite{Interleave-1F1B} divides blocks into more chunks and schedules multiple chunks launches during the warmup phase, as shown in Fig.~\ref{fig:2}(a). When the workload is divided into v chunks  (assigning layer 1, $p+1$, …, $ (v - 1)p+1$  to stage 0; layer 2, $p+2$, …, $(v-1)p+2$ to stage 1, and so forth), the bubble size is reduced to $\frac{1}{v}$ of that in the traditional 1F1B, but peak activation memory increases to $m_a(1+\frac{p-1}{pv})$ (Fig.~\ref{fig:2}(b)). Zero Bubble~\cite{ZeroBubble} splits backward propagation (BP) into activation gradient (BPA) and weight gradient (BPW), enabling a nearly 'bubble-free' pipeline. BPA-generated gradients must persist until BPW completion, which raises memory demands. Additionally, splitting the BP disrupts reuse of activation gradients for all-gather operations between BPA and BPW, and reduces the overlap between computation and communication.

\subsection{Recomputation and Offloading Strategy}\label{Recomp}
Recomputation allows only a subset of activations generated in the forward pass (called checkpoints) to be retained, with all other activations regenerated during the backward pass, trading off memory usage for additional computation. Previous work~\cite{ChainTrainingMem} introduced recomputation into DNNs, proving that optimal scheduling enables memory savings at sublinear cost for chain-structured DNN. Additionally, the benefits of recomputation vary across different operators. Recent studies suggest recomputing only the attention part in transformers~\cite{SeqParallel} or recomputing some non-linear operators~\cite{MegatronKwai} to reduce memory usage with minimal computational overhead effectively. \textbf{So far, it is still expensive to recompute the projection operator in LLM, which dominates the remaining activation memory usage}. \textcolor{black}{Moreover, ~\cite{DTR} further introduced temporal locality-aware recomputation principle and demonstrated its efficacy in data parallelism. However, temporal locality-aware recomputation in pipeline parallelism remains underexplored, necessitating systematic solutions for inter-stage dependency management and programming complexity mitigation.}

Offloading techniques allow tensors to be moved from memory-constrained GPUs to other GPUs or CPUs, trading memory usage for data movement. In BPipe~\cite{Bpipe}, GPU offloading is used to address activation memory imbalances in pipeline parallelism. However, this approach requires high-bandwidth intra-node communication links like NVLink. CPU offloading utilizes PCIe to offload activation or model states. Theoretically, CPU offloading for activation memory reduction is well-suited for long-sequence training~\cite{MegatronKwai}. However, completely offloading all the activations can lead to significant performance degradation due to limited PCIe bandwidth. For model states, previous work, such as ZeRO-Offload~\cite{ZeRO-Offload}, offloads optimizer-step to CPUs. To minimize GPU idle time caused by latency in the CPU's optimizer step, this approach imposes significant requirements on both offload bandwidth and CPU computational capacity. Overall, Recent studies still have high bandwidth requirements for the Offloading Strategy.

\subsection{Motivation 1: Temporal-Locality Matters}\label{Motivation}

\begin{figure}[htbp]
    \centerline{
    \includegraphics[width=0.48\textwidth]{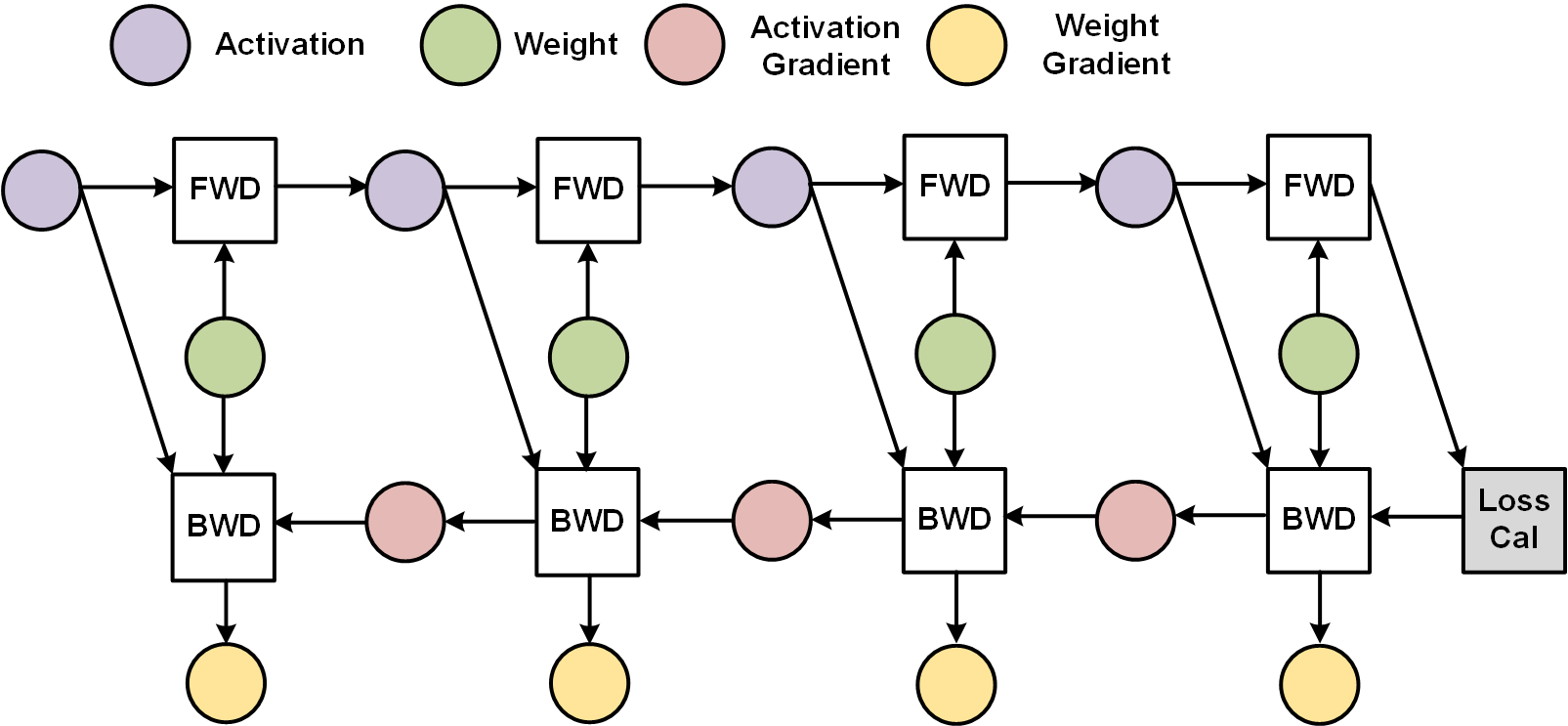}
    }
    \caption{forward and backward pass of deep neural network}
    \label{fig:3}
\end{figure}

In the training of DNN, the forward pass is conducted first, followed by backpropagation, which naturally contains differences in temporal locality. As shown in Fig.~\ref{fig:3}, activations of shallow layers are generated first during the forward pass but are the last to be released in the backward pass, resulting in poor temporal locality. In contrast, weights of deep layers can be updated early during backpropagation, though they are not needed until the end of the forward pass, providing ample time for updating the optimizer state—a property that can be further leveraged.

In pipeline parallelism, temporal locality reveals some additional interesting phenomena:

\textbf{Higher Peak Activation Storage Requirements}: The PP stage responsible for shallow-layer activations releases them last, which increases the activation storage requirements. Additionally, intervals can exist between blocks with dependencies, and these intervals can accumulate across multiple blocks, further worsening the temporal locality.

\textbf{Skewed Distribution of Peak Activation Storage}: As shown in Fig. ~\ref{fig:2}, the peak activation storage in the interleaved 1F1B tends to occur at Stage 0. When the chunk size is set to 2, the ratio of peak activation storage between shallow and deep layers is $(2p-1):p$, indicating an obvious bias.

\textbf{Sufficient Time for optimizer-step}: Beyond the natural advantage of chain-structured networks providing additional time for optimizer update on deep-layer weights, increasing the PP will provide more sufficient time for optimizer-step. Larger pipeline parallelism does not impact execution latency of each micro-batch while reducing amount of per-stage model state updates. Additionally, warm-up/cool-down phase bubbles can be strategically utilized to reduce communication and processing overhead during weight updates.

\subsection{Motivation 2: Memory-efficient PP Solutions}\label{Motivation2}

To address the memory imbalance problem in pipeline parallelism (PP), prior work primarily follow two directions. \textbf{The first way employs stage-aware strategies like AdaPipe~\cite{AdaPipe}}, which dynamically allocates varying computational workloads and recomputation ratios across pipeline stages to achieve memory balance. However, this adaptive allocation introduces significant compilation and programming complexity as it must continuously adjust to different model architectures and hardware configurations. 

\textbf{The second approach adopts locality-balanced solutions exemplified by V-Half/V-Min} ~\cite{CtrlMemPP}, which utilize a V-shaped scheduling pattern to assign tasks with the best temporal locality (deepest layer) and worst temporal locality (shallowest layer) within the same pipeline stage, thus statically balancing memory across stages while maintaining uniform computational loads.

Despite their merits, V-Half/V-Min exhibit critical limitations. First, while achieving stage-level memory balance through V-shaped scheduling, these methods still retain tensors with poor temporal locality, \textbf{failing to reach the upper bound of memory optimization via temporal locality}. Second, their memory-efficient V-shaped design requires partitioning backpropagation (BP) into separate BPA and BPW phases, preventing overlap between tensor parallelism’s collective communications and computation. \textbf{This inefficiency becomes particularly severe in long-sequence training, where it introduces persistent pipeline bubbles and degrades computational efficiency}.

\textbf{Unlike previous work, we propose TPipe, a rigorous temporal-locality-aware design to systematically address these limitations, as
detailed in the following sections}.

\begin{figure*}[htbp]
    \centerline{
    \includegraphics[width=0.85\textwidth]{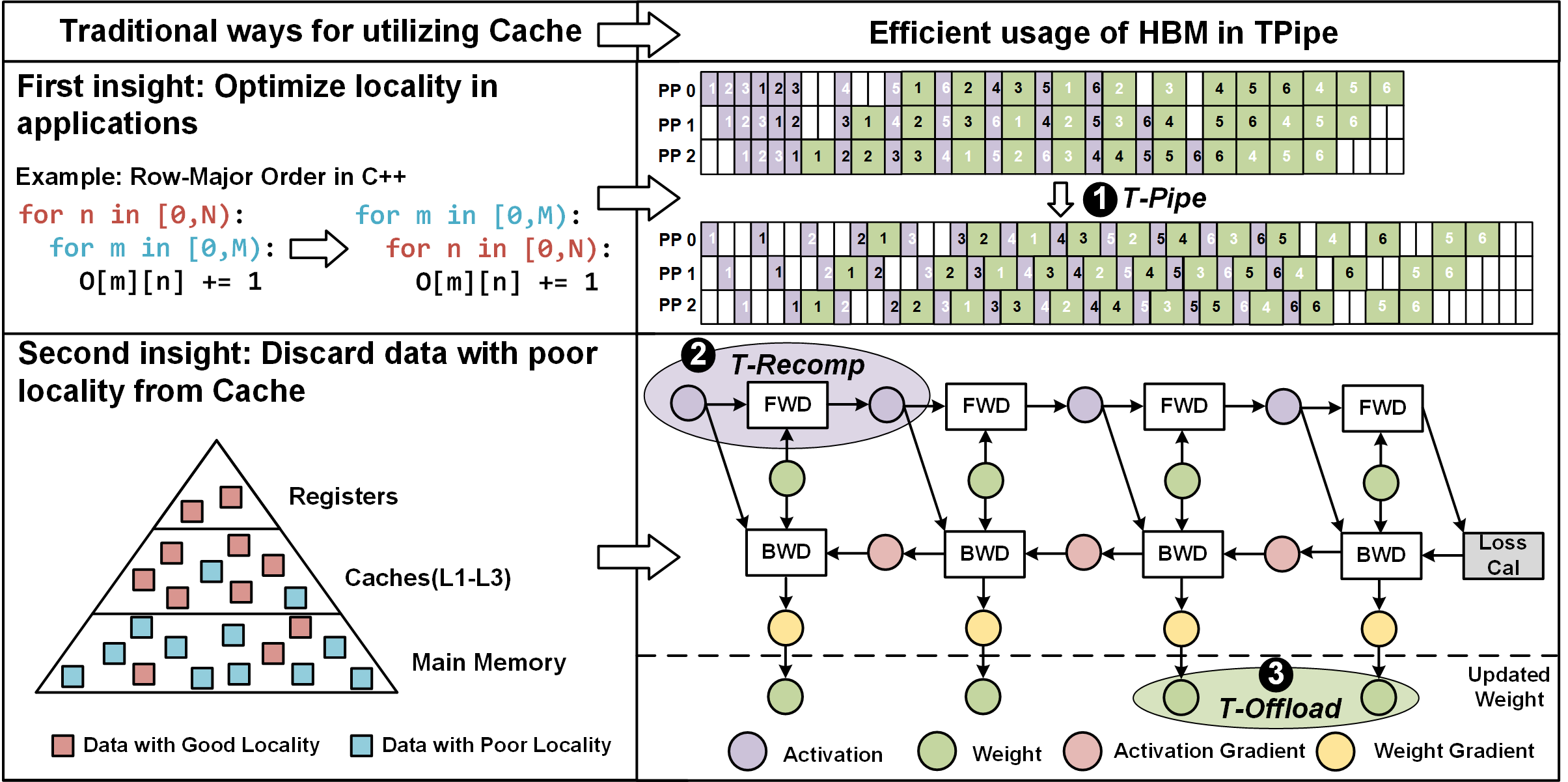}
    }
    \caption{Overview of TPipe. The core idea behind TPipe is to treat HBM as a fast but limited ”cache.” T-Pipe enhances activation temporal locality by scheduling while T-Recomp/T-Offload discard activation/weight with poor temporal locality from HBM.}
    \label{fig:4}
\end{figure*}

\section{OVERVIEW OF TPIPE}

% When the HBM storage capacity cannot fully meet the model's requirements, it can be treated as a fast but small 'cache.' The core idea behind TPipe is to optimize and exploit temporal locality during LLM pre-training to make more efficient use of this 'cache.'

Inspired by the principle of locality in cache, TPipe applies cache's core concepts to HBM. The central idea of TPipe is to optimize and leverage temporal locality during LLM training, maximizing HBM utilization efficiency.

TPipe classifies the factors that affect temporal locality in LLM training into two main types. First, temporal locality varies across different layers within the same model; this influence is referred to as \textbf{intrinsic temporal overhead}, capturing the impact on temporal locality inherently introduced by model structure. Second, the same layer may display different temporal locality depending on the scheduling strategy used; this influence is termed \textbf{extrinsic temporal overhead}, capturing the effects on temporal locality introduced by scheduling differences.

% TPipe introduces a PP scheduling strategy called T-Pipe to address extrinsic temporal overhead. As illustrated in Fig. ~\ref{fig:4}, T-Pipe’s primary goal is to advance the backward passes of both shallow and deep layers, minimizing the lifespan of activation. Detailed explanations of T-Pipe are provided in Section 4.1. This scheduling strategy reflects TPipe’s approach to optimizing temporal locality in LLM training.

\textbf{To address extrinsic temporal overhead, TPipe draws on the first critical insight from cache: optimize locality in applications}. As illustrated in Fig. ~\ref{fig:4}, TPipe introduces a PP scheduling strategy called T-Pipe to advance backward passes across all layers, minimizing the lifespan of activation. Detailed explanations of T-Pipe are provided in Section \ref{T-Pipe}. This scheduling strategy reflects TPipe’s approach to optimizing temporal locality in LLM training.

\textbf{While scheduling can only reduce extrinsic temporal overhead, addressing intrinsic temporal overhead still requires additional strategies. To tackle this problem, TPipe draws on a second critical insight from cache: discard data with poor locality from cache.} As illustrated in Fig. ~\ref{fig:4}, TPipe removes data with poor intrinsic temporal locality from HBM. This involves T-Recomp, a recomputation strategy for shallow-layer activations, and T-Offload, an offloading strategy at bubble for deep-layer weights. Detailed explanations of T-Recomp and T-Offload are provided in Section \ref{T-Recomp} and Section \ref{T-Offload}, respectively. Through recomputation and offloading, TPipe leverages temporal locality to improve memory efficiency in LLM training.

\section{TEMPORAL-LOCALITY-AWARE SCHEDULE FOR ACTIVATION}

\subsection{T-Pipe: Temporal-Locality-Aware Pipeline}\label{T-Pipe}
\begin{figure}[htbp]
    \centerline{
    \includegraphics[width=0.48\textwidth]{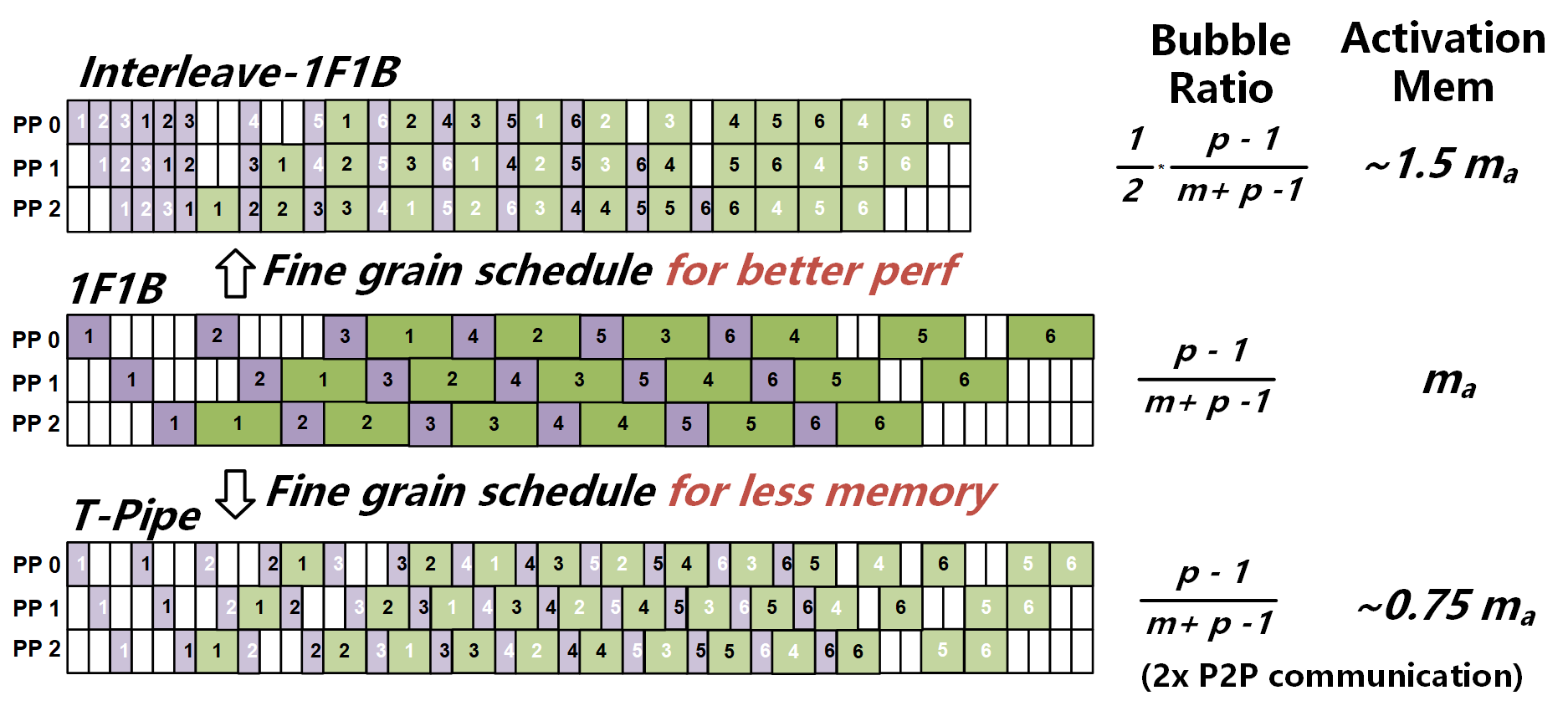}
    }
    \caption{Comparison of Interleave-1F1B, 1F1B, and T-Pipe.}
    \label{fig:5}
\end{figure}
\paragraph{\textbf{T-Pipe Schedule}}
We first analyze memory usage in 1F1B and Interleave-1F1B through intrinsic and extrinsic temporal overhead perspectives. As Fig. ~\ref{fig:5} illustrates, while different DNN layers exhibit varying temporal locality, \textbf{1F1B's coarse-grained execution fails to leverage intrinsic locality differences among layers assigned to the same stage. Although Interleave-1F1B employs chunk-based fine-grained pipelining to reveal intrinsic locality variations within stages, it incurs higher activation memory than 1F1B due to significant extrinsic temporal overhead}. For example, the extrinsic overhead caused by accumulated intervals during the computation of chunk 2 (micro-batch 4) equals the intrinsic overhead required for DNN computation.

Therefore, T-Pipe's core innovation lies in reducing extrinsic temporal overhead while leveraging intrinsic locality variations through fine-grained scheduling. Similar to Interleave-1F1B, T-Pipe partitions workloads into chunks to expose intrinsic locality variations within stages (Fig. ~\ref{fig:5}). However, it uniquely prioritizes early backpropagation completion to minimize extrinsic overhead. 

Interleave-1F1B and T-Pipe can be viewed as dual optimization approaches: while both employ chunk-based fine-grained pipelining, the former prioritizes computational performance, whereas the latter emphasizes memory efficiency. Guided by extrinsic overhead minimization principles, T-Pipe achieves a peak activation memory of just 75\% $m_a$, halving Interleave-1F1B's activation storage requirements. This memory advantage comes with a fundamental trade-off: The bubble ratio of T-Pipe doubles that of Interleave-1F1B, similar to 1F1B. These quantitative results are rigorously derived from the theoretical analysis presented in Appendix \ref{Appendix T-Pipe}.

Driven by intrinsic temporal locality variations, shallow-layer activations account for 50\% $m_a$, while deep-layer activations contribute 25\% $m_a$. We will implement targeted technique to exploit this fundamental temporal disparity, as detailed in the subsequent section.

% \paragraph{\textbf{Implementation of T-Pipe}} 

% In T-Pipe’s implementation, partial asynchronous communication is introduced to reduce bubble size. In mainstream frameworks like DeepSpeed and Megatron, the 1F1B configuration uses synchronous P2P communication. However, in T-Pipe, there can be significant intervals between chunks during forward and backward passes. If synchronous P2P communication is used in these intervals, much of the time would be spent waiting for communication to complete. To address this, T-Pipe modifies P2P communication between chunks to be asynchronous, thereby achieving higher throughput. As illustrated in Fig. ~\ref{fig:5}(b), after Stage 0 completes the backward computation for chunk 2, it can proceed with subsequent tasks until just before Stage (P - 1) begins the corresponding backward pass for chunk 1, at which point the P2P communication is completed. Other P2P communications remain synchronous, which is consistent with mainstream frameworks.

\begin{figure*}[htbp]
    \centerline{
    \includegraphics[width=0.95\textwidth]{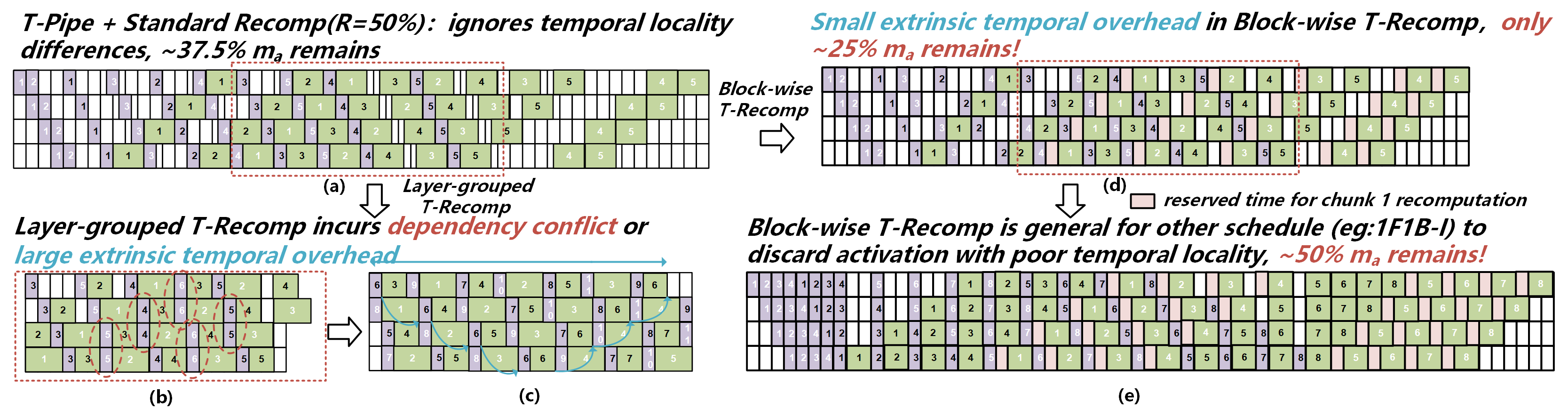}
    }
    \caption{(a) T-Pipe + Standard Recomp(R=50\%). (b) Dependency conflict would occur in T-Pipe+layer-grouped T-Recomp. (c) removing dependency conflict in T-Pipe+layer-grouped T-Recomp results in large extrinsic temporal overhead. (d) T-Pipe+block-wise T-Recomp. (e) Interleave-1F1B+block-wise T-Recomp.}
    \label{fig:6}
\end{figure*}

\subsection{T-Recomp: Temporal-Locality-Aware Recomputation}\label{T-Recomp}
\paragraph{\textbf{Why T-Recomp}}

T-Pipe successfully minimizes extrinsic temporal overhead, but remains fundamentally constrained by the intrinsic temporal locality differences in the chain structure of DNN.

To further improve memory efficiency, we can evict data with poor temporal locality from HBM. This aligns with fundamental principle in computer architecture: caches with limited capacity should retain only data exhibiting good locality. We systematically adapt this well-established caching paradigm to HBM management.

To address eviction of low-temporal-locality data from HBM, the most intuitive approach involves offloading target activations to storage in next level (e.g., CPU DRAM). However, this solution demonstrates critical limitations: its performance becomes strictly bandwidth-bound due to repeated offload/reload cycles across all training batches, each requiring identical data movement operations. Given that compute scaling has consistently outpaced memory and communication bandwidth scaling historically, this approach is nonuniversal: In systems exhibiting significant compute-to-bandwidth asymmetries (e.g., wafer-scale integrated system~\cite{Dojo}~\cite{Cerebras} or PCIe-bandwidth-constrained GPU clusters), it incurs severe throughput degradation.

A more viable solution involves directly evicting low-temporal-locality data from the memory hierarchy and recomputing it when required. Given that computation scaling consistently outpaces memory and communication bandwidth scaling in hardware evolution, rendering computational resources relatively inexpensive, computation emerges as a hardware-agnostic technique without presupposing specific memory or interconnect bandwidth configurations. We therefore employ recomputation to evict activation values with poor temporal locality from HBM. \textbf{This approach strategically prioritizes recomputation of shallower layers, a methodology we designate as T-Recomp}.

\paragraph{\textbf{Concerns Regarding T-Recomp}}

T-Recomp fundamentally differs from conventional recomputation in pipeline parallelism. Traditional coarse-grained PP fails to reveal temporal locality variations within single stage, necessitating T-Recomp's implementation in multi-chunk fine-grained pipelining. \textbf{Under standard recomputation, all chunks perform recomputation with same ratio, resulting in uniform backpropagation times across chunks. However, T-Recomp strategically prioritizes recomputation of shallower-layer activations, intentionally creating nonuniform backpropagation times between chunks}, a key innovation that aligns recomputation with intrinsic temporal locality patterns.

Conventional recomputation employs a layer-grouped strategy for memory efficiency, partitioning transformer layers within each chunk into discrete groups while preserving only input activation of each group. During recomputation, buffer must be allocated for regenerated activation tensors. To minimize buffer capacity, recomputation is triggered just-in-time when activations of specific group are required. This design prevents accumulation of recomputed results, which has become prevalent in standard implementations.

Unfortunately, the layer-grouped recomputation approach is incompatible with T-Recomp. \textbf{This method forces subsequent computations within the same stage to wait not only for the current recomputation but also for accumulated delays from preceding stages.} While conventional recomputation maintains uniform backpropagation times across chunks, preserving consistent inter-stage dependencies, T-Recomp's variable recomputation times (as shown in Fig. ~\ref{fig:6}(b)) disrupt this balance. \textbf{The resulting accumulation of recomputation alters inter-stage dependency patterns}, creating dependency conflicts. Attempting to resolve these violations through schedule adjustments like advancing conflicting tasks as illustrated in Fig.~\ref{fig:6}(c) introduces substantial extrinsic temporal overhead, ultimately undermining motivation for temporal locality optimization.

\paragraph{\textbf{Addressing the Concerns}} 

The key solution lies in preventing accumulated recomputation from affecting dependency relationships, which we achieve through block-wise T-Recomp. \textbf{As illustrated in Fig.~\ref{fig:6}(d), we decouple recomputation from the dependent backpropagation process, thereby eliminating the negative impact of recomputation accumulation on inter-stage dependencies}. 

While block-wise T-Recomp successfully prevents dependency conflicts within chunks, inter-chunk dependencies remain critical. As illustrated in Fig.~\ref{fig:6}(d), block-wise T-Recomp introduces intervals during forward computation. As pipeline stage (P) increases, these intervals accumulate—progressively shortening the time between the completion of chunk 1’s forward pass and the initiation of chunk 2’s forward pass, eventually triggering inter-chunk dependency conflicts.

Inter-chunk dependency conflicts can be effectively resolved. When P increases (e.g., $P\geq8$), delaying chunk 2's forward pass effectively prevents inter-chunk dependency conflicts. Additionally, we demonstrate in Appendix \ref{Appendix A} that for $P\leq40$, delaying the computation of chunk 2 by just one round is sufficient to avoid dependency conflicts between chunks---a parallelism level that significantly exceeds typical configurations. Notably, chunk 2's forward pass starts later but completes its backward pass earlier than chunk 1, ensuring no impact on total runtime.

\paragraph{\textbf{Benefits of T-Recomp}} 
T-recomp demonstrates better recomputation efficiency. We can see that block-wise recomputation introduces some intervals during forward computation, resulting in a slight increase in extrinsic temporal overhead. However, because recomputation proportionally extends the execution cycles of workloads in pipeline parallelism, it does not increase storage requirements for deep-layer activations under T-Recomp. While this approach requires larger buffer compared to layer-grouped recomputation, it does not significantly impact overall activation storage efficiency. These findings are rigorously analyzed in Appendix \ref{Appendix T-Recomp}. In total, the combined storage footprint of activations and buffers is approximately 25\% $m_a$—\textbf{only one-fourth of the activation storage required by standard 1F1B and a 2× reduction compared to 1F1B+50\% recomputation}. This demonstrates T-Pipe+T-Recomp's superior memory efficiency.

T-Recomp demonstrates broader applicability beyond T-Pipe. Since its block-wise recomputation design preserves dependency relationships intact, any pipeline parallelism approach exhibiting intra-stage temporal locality variations can effectively adopt T-Recomp. Fig.~\ref{fig:6}(e) illustrates this generalization through T-Recomp-enhanced Interleave-1F1B scheduling: after recomputing shallow-layer activations with poor temporal locality, the system retains only $\sim$50\% $m_a$. This enables Interleave-1F1B+T-Recomp to match the memory footprint of 1F1B+50\% standard recomputation while halving the bubble compared to 1F1B+50\% standard recomputation.

\section{TEMPORAL-LOCALITY-AWARE SCHEDULE FOR MODEL STATE}\label{T-Offload}
% Temporal locality can be applied not only to activation memory usage optimization but also to model state store optimization.

\paragraph{\textbf{T-Offload Scheduling Principle}}

As outlined in Motivation 1, model states exhibit temporal locality variations across layers similar to activations, enabling prioritized eviction of poor locality model states from HBM. Unlike activations that can be recomputed, model states require offloading to DRAM when evicted, an approach we term T-Offload. But offloading model states only happens once per mini-batch, resulting in significantly lower data movement volume. Quantitative analysis using Llama3's technical report data~\cite{LLAMA3} reveals this disparity: for a batch size of 16M tokens, model state size of Llama3-405B is approximately three orders of magnitude smaller than its activation memory footprint.

\paragraph{\textbf{Hardware-agnostic T-Offload Solution by Co-design with T-Pipe}}

\begin{figure}[htbp]
    \centerline{
    \includegraphics[width=0.48\textwidth]{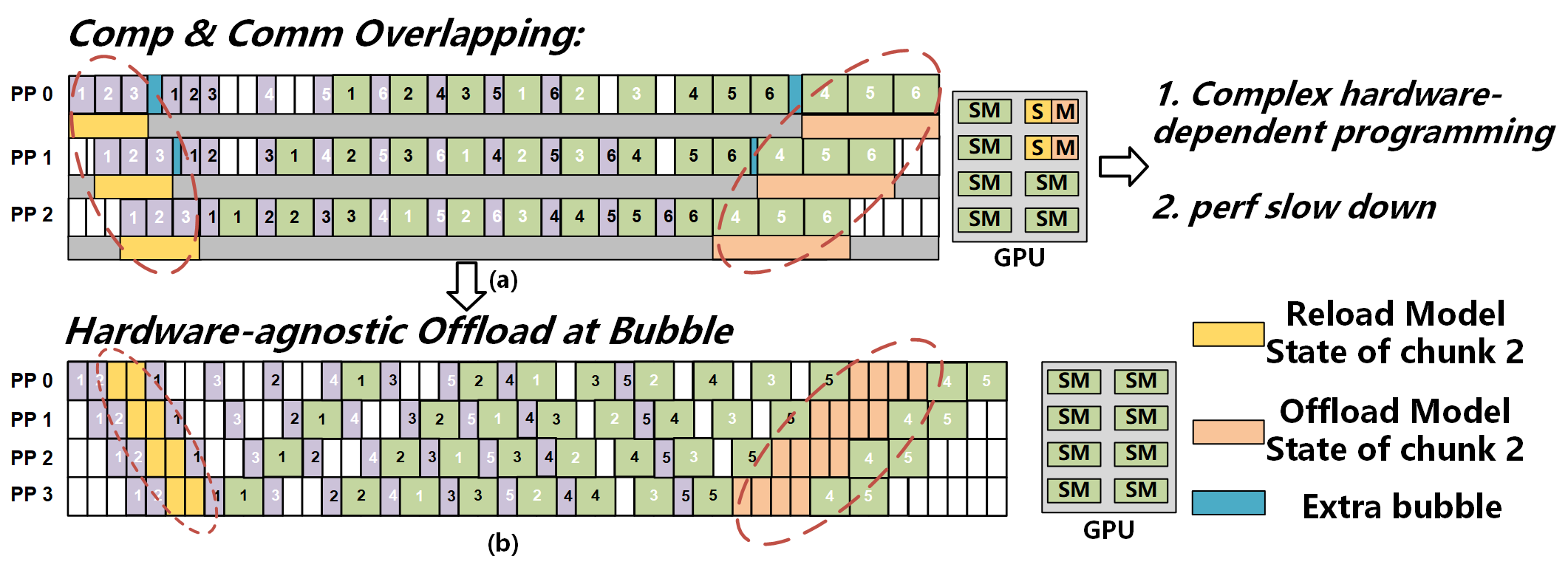}
    }
    \caption{(a)T-Offload in form of overlapping. (b)hardware-agnostic T-Offload in T-Pipe.}
    \label{fig:8_T-Offload}
\end{figure}

However, implementing T-Offload presents varying challenges across different pipeline parallelism (PP) strategies. In existing PP, the communication required for T-Offload overlaps with warm-up/cooldown computations, introducing non-trivial overhead. As shown in Fig.~\ref{fig:8_T-Offload}(a), the communication tasks consume a portion of Streaming Multiprocessor (SM) resources on GPUs, which not only increases programming complexity but also degrades computational performance. Specifically, competition for SM resources slows down the execution of overlapping compute blocks, and this slowdown introduces additional bubbles during the warm-up and cooldown phases. These issues become pronounced in hybrid DP-PP-TP scenarios. While dedicated asynchronous communication engines could potentially address these issues, such solutions remain hardware-specific and introduce their own programming complexity.

In contrast, T-Pipe offers a hardware-agnostic and simple integration solution. As illustrated in Fig.~\ref{fig:8_T-Offload}(b), T-Pipe naturally introduce bubbles that can be utilized by T-Offload. This phenomenon stems from T-Pipe's core optimization for activation temporal locality: by eliminating intervals between pipeline tasks to maximize locality, the inherent dependencies between chunks automatically generate these bubbles during warm-up and cooldown phases.

These bubbles enable straightforward T-Offload implementation during idle phases, as shown in Fig.~\ref{fig:8_T-Offload}(b). During the cooldown phase, we schedule offloading chunk 2’s gradients and performing its optimizer updates between chunk 2’s backward pass completion and chunk 1’s backward pass initiation. Quantized weight reloads are allocated to the idle time between two chunk forward computations in the next mini-batch’s warm-up phase. This approach preserves throughput with simplicity, avoiding the programming overhead and resource contention inherent to computation-communication overlap—a key advantage over complex overlap-dependent methods. \textbf{Notably, T-Pipe’s bubble exhibits linear growth with pipeline stage (P), confirming that this hardware-friendly T-Offload approach retains scalability}. Our rigorous analysis in Appendix ~\ref{Appendix T-Offload} demonstrates that the solution maintains robust scaling with PP,DP and sequence length. Furthermore, the optimizer step involves synchronized operations, such as gradient norm calculations, which aggregate all gradients. Since T-Offload splits optimizer updates between the CPU and GPU, synchronization overhead must be addressed. To mitigate this, a post-optimizer update validation strategy, similar to~\cite{ZeroBubble}, can be employed to manage these challenges effectively.

\paragraph{\textbf{Compatibility of T-Offload}}

% T-Offload is designed to be compatible with a variety of scheduling strategies. Our approach exclusively leverages the 'bubbles' created within T-Pipe, eliminating competition over communication bandwidth and CPU with other scheduling approaches. As a result, it seamlessly supports many parallelism strategies, including tensor parallelism, context parallelism, and data parallelism. This flexibility makes it widely applicable across diverse distributed parallelism training frameworks.

T-Offload operates independently of scheduling strategies by utilizing T-Pipe's intrinsic idle-phase bubbles for offloading operations. This enables native support for tensor, context, and data parallelism, ensuring broad applicability across distributed training environments.

% T-Offload is also fully compatible with the CPU Activation Offloading scheme. It performs Model State Offloading in the interval between the completion of chunk 2’s backward computation and the onset of the forward computation for the next mini-batch of chunk 2. This process exclusively utilizes the PCIe upload bandwidth during the warm-up phase and the PCIe offload bandwidth during the cool-down phase. This approach does not interfere with CPU Activation Offloading; rather, it enhances the efficient utilization of the PCIe bidirectional bandwidth.

T-Offload seamlessly integrates with CPU activation Offloading. It offloads model states during bubbles, leveraging PCIe upload bandwidth during warm-up and offload bandwidth during cool-down. This strategy complements activation Offloading by maximizing PCIe bidirectional bandwidth utilization without contention.

Therefore, T-Offloading effectively utilizes the temporal locality of weights while also maximizing the use of the prevalent 'bubbles' during pipeline parallelism's warm-up and cool-down phases. Consequently, it can seamlessly integrate with numerous scheduling optimization strategies.

\begin{figure*}[htbp]
    \centerline{
    \includegraphics[width=0.90\textwidth]{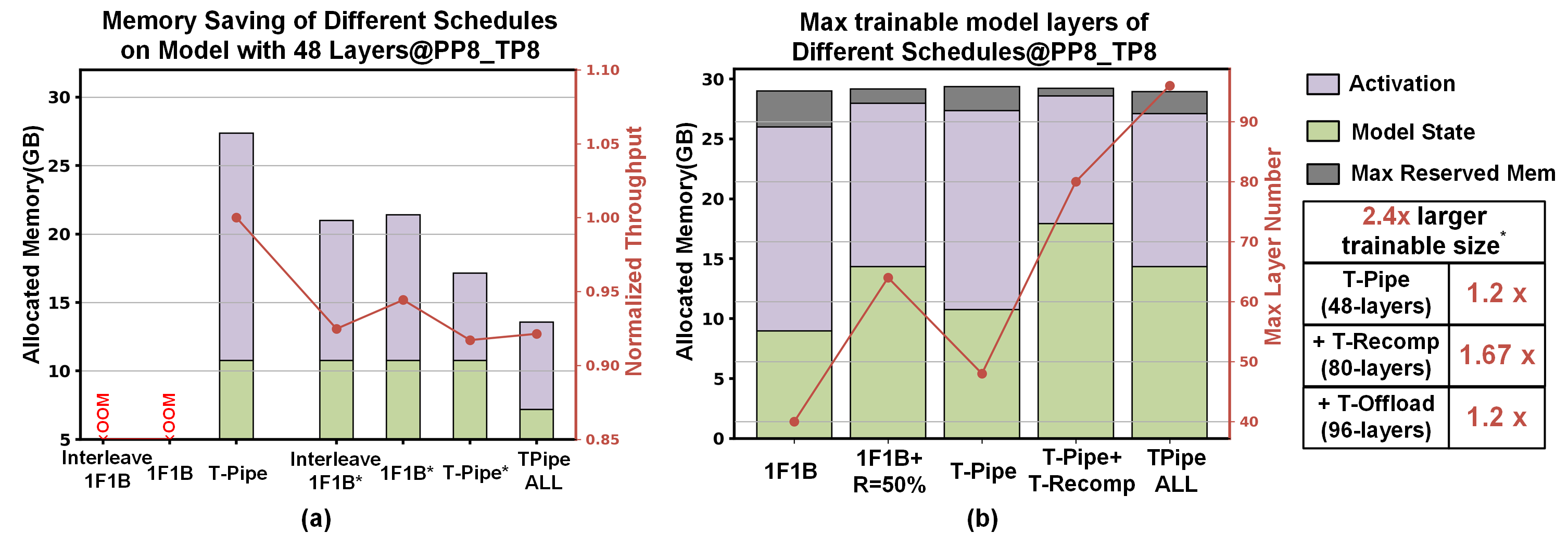}
    }
    \caption{End-to-End Evaluation on TPipe. Input of ColumnLinear is not evenly divided along the TP dimension for better performance. (a) shows the memory saving of different schedules on a 48-layer(LLAMA 42B) model at (PP, TP)=(8,8), global batch=128, micro batch=2, seq len=4K. *: R=66.7\% in interleave 1F1B, R=50\% in 1F1B and T-Recomp is used in T-Pipe. (b) illustrates the maximum trainable LLAMA-based model size of different schedules at (PP, TP)=(8,8), global batch=128, micro batch=2, seq len=4K. *: compared with 1F1B.}
    \label{fig:8}
\end{figure*}

\section{EXPERIMENTAL RESULTS}

This section addresses three critical research questions: \\
\noindent
$\bullet$ How does TPipe perform in terms of memory efficiency and throughput (Section ~\ref{GPU_ecaluation})? \\
$\bullet$ What are TPipe's advantages and limitations against other memory-efficient pipeline parallelism approaches (Section ~\ref{Sim_ecaluation})? \\
$\bullet$ What potential gains could be achieved through finer-grained TPipe optimizations (Section ~\ref{DSE_ecaluation})?

\subsection{Methodology}

\textcolor{black}{\textbf{Setup.} Our experiments were carried out by default on a cluster comprising up to 64 GPUs, organized across 8 nodes. Nodes are interconnected with four 200 Gbps NICs. Each node is equipped with 8 GPUs and 2 CPUs(Intel Xeon Platinum 8480+) and 2TB DRAM, 15.36TB NVME. Each GPU features 32GB of HBM and connects to the CPU through PCIe5 x8(32GB/s).}

To thoroughly evaluate the benefits of temporal locality optimization, our experiments require extensive configuration tuning for design space exploration. To accelerate this process, we have augmented our experimental validation with large-scale A100 cluster simulations. Each GPU in the A100 cluster is equipped with 80GB of HBM memory, with 300GB/s NVLink intra-node connections and 100Gbps inter-node scale-out bandwidth per GPU. For performance evaluation, we profile operator execution times while accounting for collective communication overhead, deriving pipeline stage durations and quantify pipeline bubbles and P2P communication using established methods ~\cite{SimAI}. Memory requirements become analytically tractable once scheduling strategies are fixed. This simulation framework enables precise comparison between TPipe and other SOTA solutions.

\textcolor{black}{\textbf{Workloads.}} We evaluate TPipe on two representative models—LLAMA2-70B~\cite{LLAMA2} and GPT3-175B~\cite{GPT3}. Across different experiments, the sequence length varies and model size is adjusted by modifying the number of model layers.

\textcolor{black}{\textbf{Baselines.}} TPipe is benchmarked against 1F1B and 1F1B-I(Interleave-1F1B) to quantify the benefits of the temporal locality. Recomputation-enabled variants are labeled as "strategy+R=X\%", e.g., 1F1B+R=50\% denotes 50\% recomputation.

Additionally, through simulations, we comprehensively compare TPipe with other memory-efficient pipeline parallelism strategies, including AdaPipe (which employs stage-aware task scheduling)~\cite{AdaPipe} as well as V-Min and V-Half (which utilize locality-balanced task scheduling)~\cite{CtrlMemPP}. When analyzing AdaPipe, we set the same storage budget for AdaPipe and T-Pipe + T-Recomp and compared the differences in their MFU (Model FLOPs Utilization) to conduct a more comprehensive and fair comparison.

\textcolor{black}{For the following experiments, all evaluations were performed on real GPU clusters unless otherwise indicated. Moreover, operator-level recomputation~\cite{MegatronKwai} and FlashAttention~\cite{Flashattention} are enabled by default.}

\subsection{End-to-End Evaluation}\label{GPU_ecaluation}

Fig. ~\ref{fig:8}(a) demonstrates TPipe’s advantage in HBM demand reduction and throughput preservation. Under PP8\_TP8, TPipe trains a 48-layer (LLAMA 42B) model where Interleave-1F1B/1F1B trigger OOM. T-Recomp slashes activation memory from 16.61GB to 6.39GB, outperforming 1F1B+R=50\% (10.65GB activation usage is needed). With T-Offload, TPipe ALL achieves 97.58\% throughput of 1F1B+R=50\% while using only 63.35\% of activation and model state memory.

Moreover, fig. ~\ref{fig:8}(b) quantifies TPipe's memory efficiency under PP8\_TP8 (32GB HBM): While 1F1B supports only 40 layers (requiring 50\% recomputation for 64 layers), T-Pipe trains 48-layer models natively, extendable to 80 layers via T-Recomp and 96 layers (LLAMA 84B) with further T-Offload integration—all while keeping the max reserved memory at roughly the same level.This achieves 2.4× and 1.5× model scale improvements over 1F1B and 1F1B+R=50\% respectively, demonstrating temporal locality's critical role in PP.

\begin{figure}[htbp]
    \centerline{
    \includegraphics[width=0.40\textwidth]{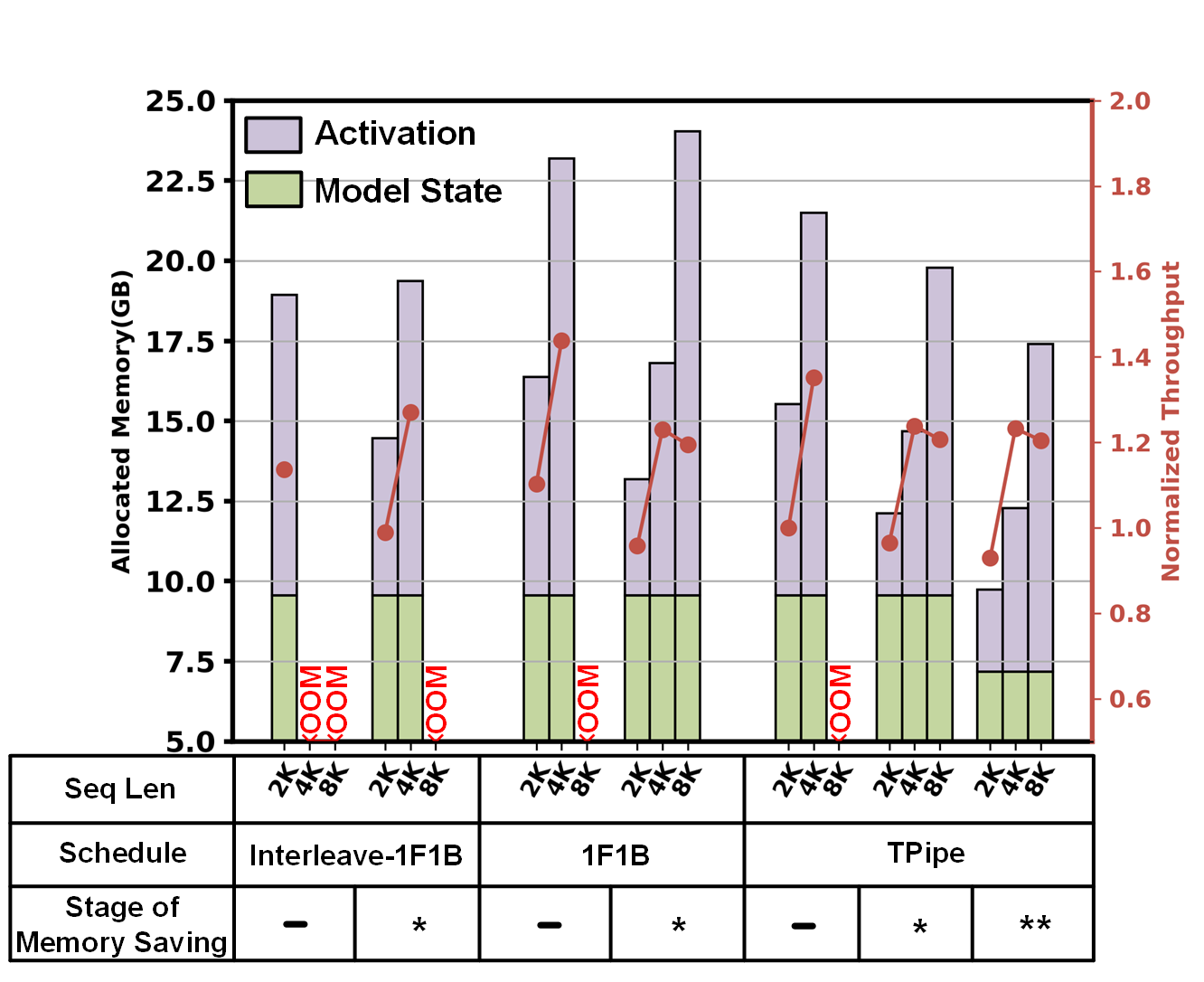}
    }
    \caption{memory and throughput of different schedules under various sequence lengths on a 32-layer LLAMA-based model at (DP, PP, TP)=(2,4,8), global batch=128, micro batch=2. *: R=50\%, T-Recomp is used in TPipe. **: T-Offload is further adopted. Input of ColumnLinear is not evenly divided along the TP dimension for better performance.}
    \label{fig:9}
\end{figure}

We further study the behavior of different scheduling strategies across various sequence lengths, shown in Fig.~\ref{fig:9}. Under the DP2 (ZeRO-1)\_PP4\_TP8 configuration, Interleave-1F1B exhibits the worst memory efficiency, running out of memory (OOM) at a sequence length of 8k, even with R=50\%. Theoretical analysis confirms smaller PP degrades TPipe’s memory gains—achieving only 12.5\% (T-Pipe) and 25\% (T-Recomp) activation memory reduction versus 1F1B and its variants, but these savings scale markedly with longer sequences. T-Offload further reduces model state memory, enabling TPipe to sustain efficiency as both sequence length and model size increase. In terms of throughput, while T-Pipe experiences a relative throughput drop of 6\%–9\% compared to 1F1B, both T-Recomp, and TPipe ALL maintain throughput comparable to 1F1B+R=50\% across all sequence lengths despite minor throughput fluctuations from the Operator Library.

\subsection{Analysis about Memory-efficient PP Solutions}\label{Sim_ecaluation}

\begin{figure}[htbp]
    \centerline{
    \includegraphics[width=0.48\textwidth]
    {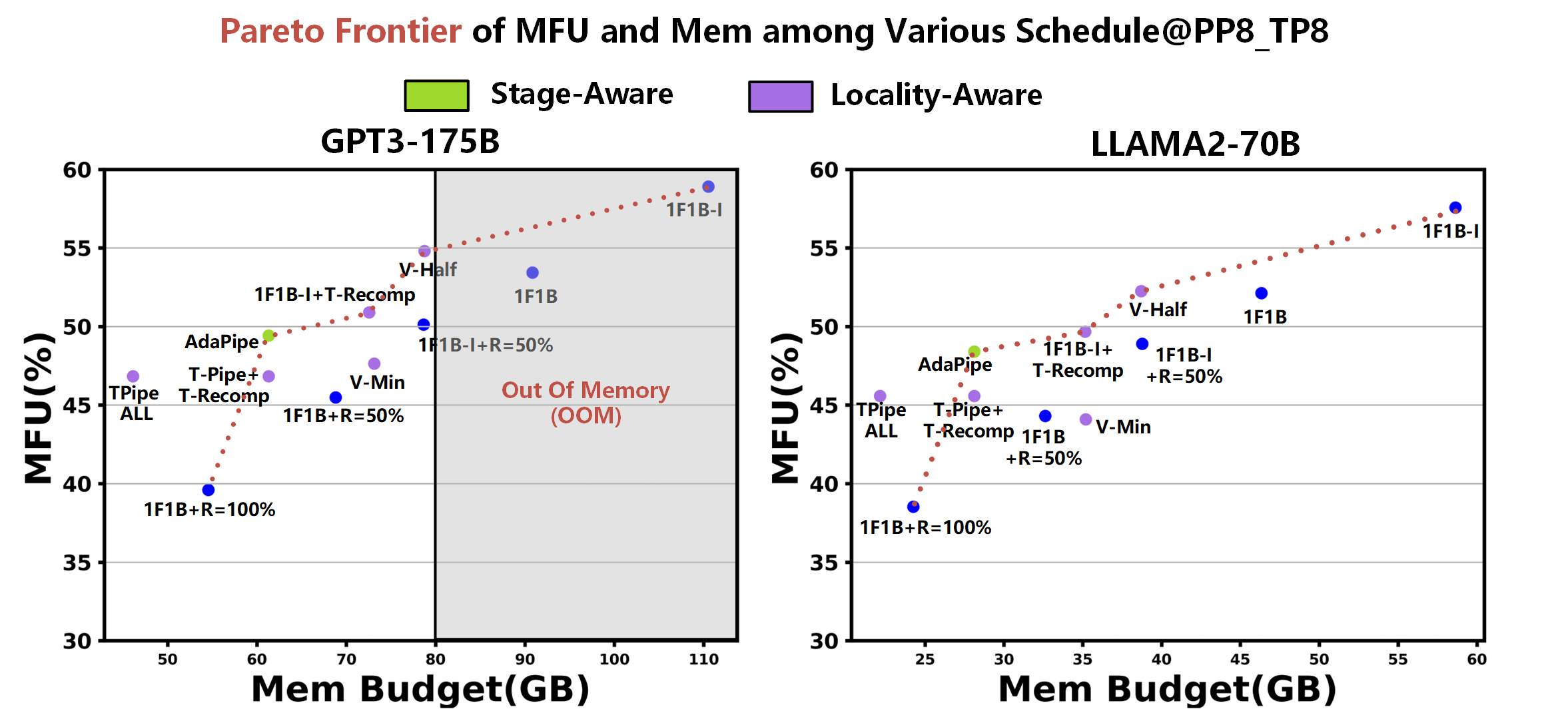}
    }
    \caption{Pareto Frontier of MFU and Mem among Various Schedule@PP8\_TP8 A100 cluster. Global batch=32,micro batch=1, seq len=16K.}
    \label{fig:R1}
\end{figure}

Fig. ~\ref{fig:R1} presents the MFU and memory characteristics across different scheduling strategies, where memory-efficient pipeline parallelism (PP) approaches dominate the Pareto frontier. Compared to memory-optimized V-Min and 1F1B+R=50\%, T-Pipe+T-Recomp achieves 19.27\%-25.14\% and 12.23\%-16\% higher memory efficiency respectively while maintaining comparable or superior MFU. When evaluated against AdaPipe, T-Pipe+T-Recomp replaces complex dynamic compilation and search mechanisms with a simpler static approach, incurring only a 2.59\%-2.82\% MFU reduction at equivalent memory usage. The solution further demonstrates seamless integration with T-Offload, where TPipe ALL delivers 26.98\%-32.92\% memory efficiency gains over AdaPipe, establishing itself as both simpler and more memory-efficient. Under relaxed memory constraints, Interleave-1F1B(1F1B-I) with T-Recomp achieves higher MFU than the basic R=50\% scheme while saving approximately 25\% activation storage, with both Interleave-1F1B(1F1B-I)+T-Recomp and V-Half exhibiting optimal MFU-memory trade-offs that define the Pareto frontier.

\begin{figure}[htbp]
    \centerline{
    \includegraphics[width=0.48\textwidth]
    {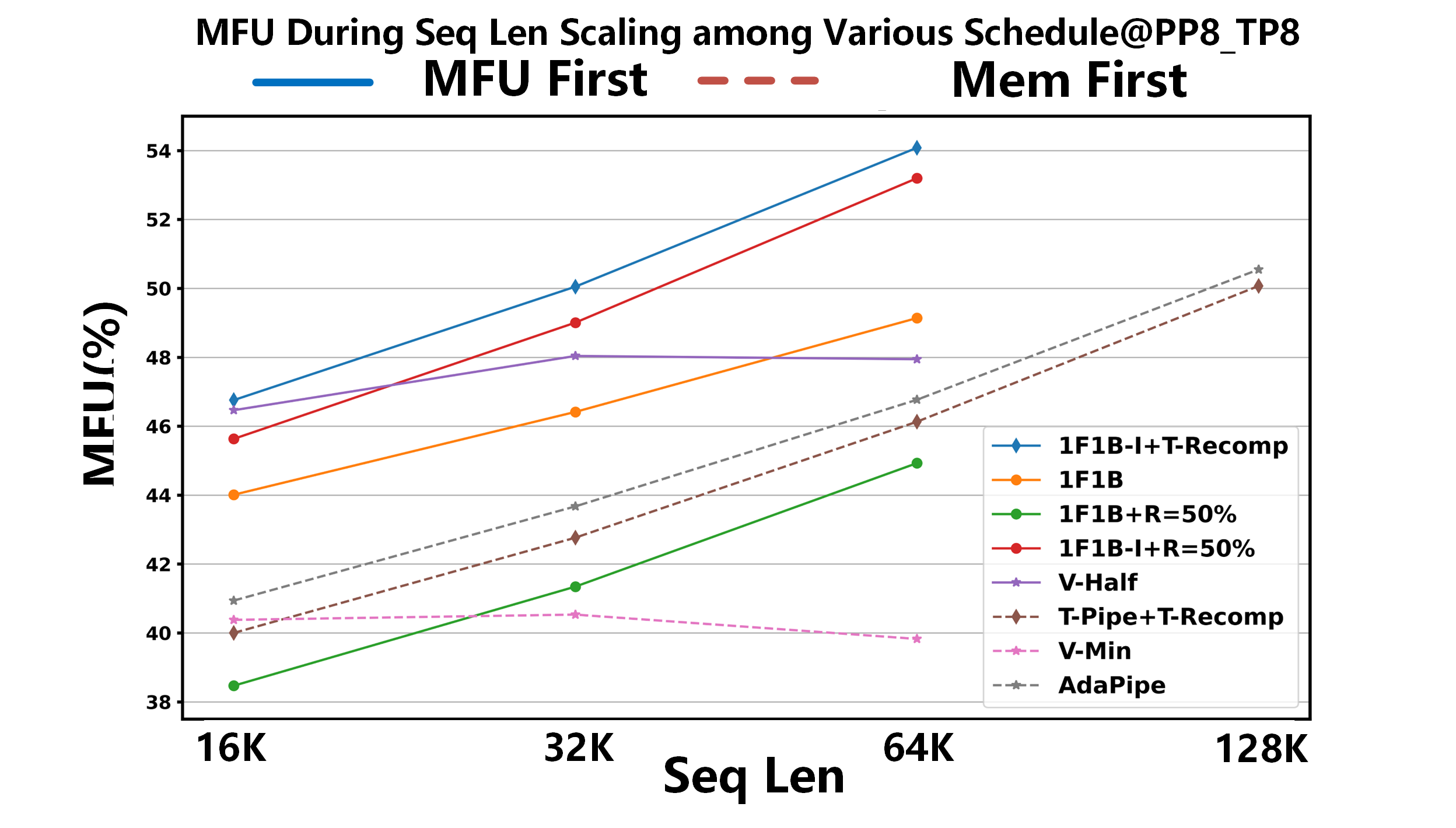}
    }
    \caption{MFU of different PP schedule on 48 layers LLAMA-based model@PP8\_TP8 A100 cluster when scaling context length. global batch=16, micro batch=1. \textit{\textbf{Neither T-Recomp nor R=50\% solution incorporates recomputation for attention mechanisms for better MFU.}}}
    \label{fig:LongSeq_1}
\end{figure}

\begin{figure}[htbp]
    \centerline{
    \includegraphics[width=0.48\textwidth]
    {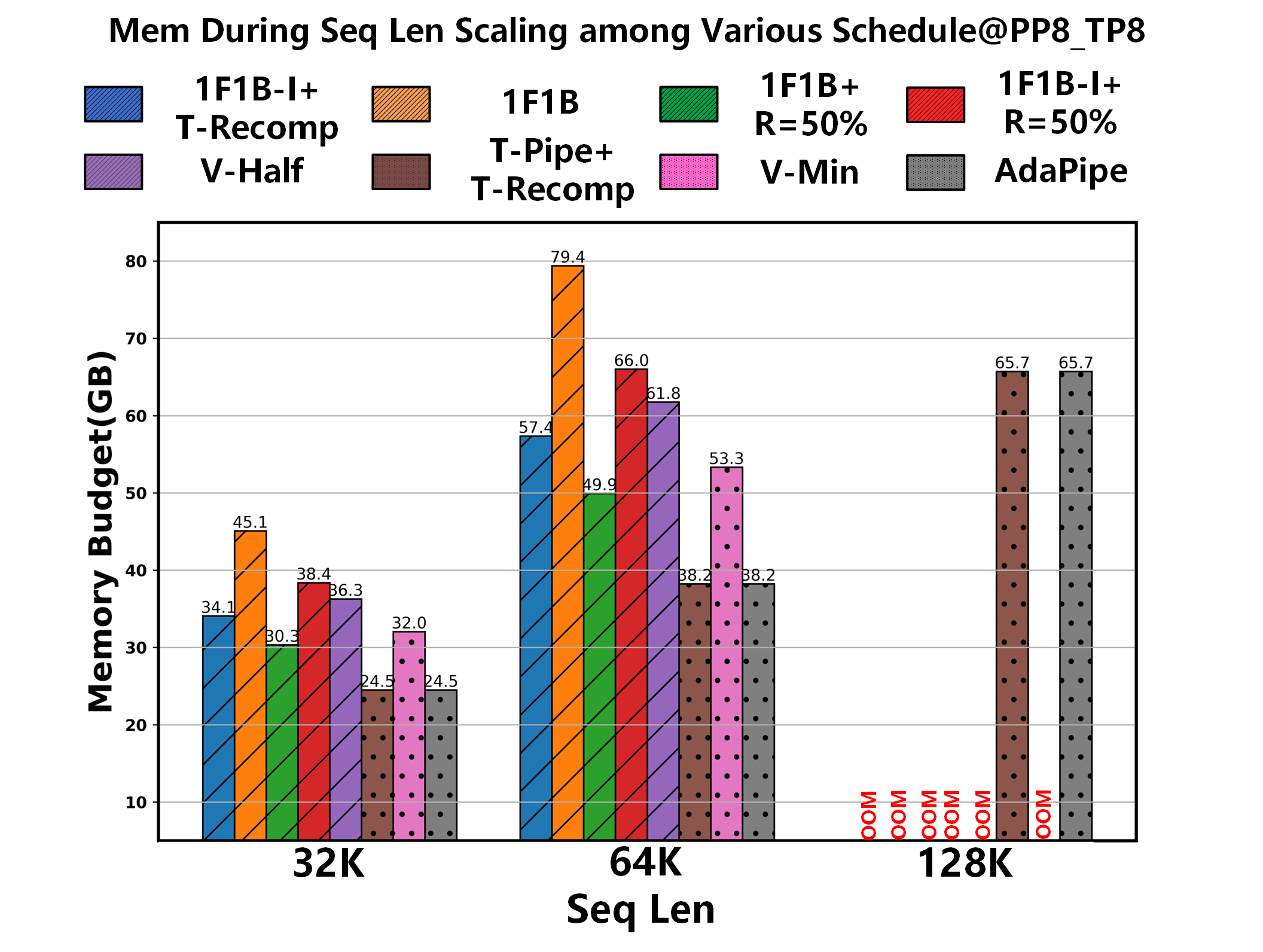}
    }
    \caption{Memory budget of different PP schedule on 48 layers LLAMA-based model@PP8\_TP8 A100 cluster when scaling context length. global batch=16, micro batch=1. \textit{\textbf{Neither T-Recomp nor R=50\% solution incorporates recomputation for attention mechanisms for better MFU.}}}
    \label{fig:LongSeq_2}
\end{figure}

Fig. ~\ref{fig:LongSeq_1} and ~\ref{fig:LongSeq_2} present the MFU and memory budget under long-sequence training scenarios, where TPipe demonstrates progressively greater computational and memory advantages as sequence length increases. Compared to standard 1F1B, the 1F1B-I+T-Recomp configuration achieves 24.58\%-38.41\% higher memory efficiency while simultaneously improving MFU by 2.75\%-4.94\%. Although V-Half maintains comparable memory efficiency to 1F1B-I+T-Recomp, their MFU gap widens significantly with longer sequences - reaching a 6.14\% difference (47.94\% vs 54.08\%) at 64K sequence length. This divergence stems from V-Half's partitioned backpropagation (BPA and BPW) which introduces increased pipeline bubbles during long-sequence training. For similar reasons, T-Pipe+T-Recomp outperforms V-Min by 6.30\% MFU (46.13\% vs 39.83\%) at 64K while maintaining superior memory efficiency. Furthermore, as sequences extend from 16K to 128K, the MFU gap between T-Pipe+T-Recomp and AdaPipe narrows from 0.94\% to 0.48\% under equivalent memory budgets, demonstrating comparable computational efficiency with significantly simpler implementation.

\begin{figure}[htbp]
    \centerline{
    \includegraphics[width=0.48\textwidth]
    {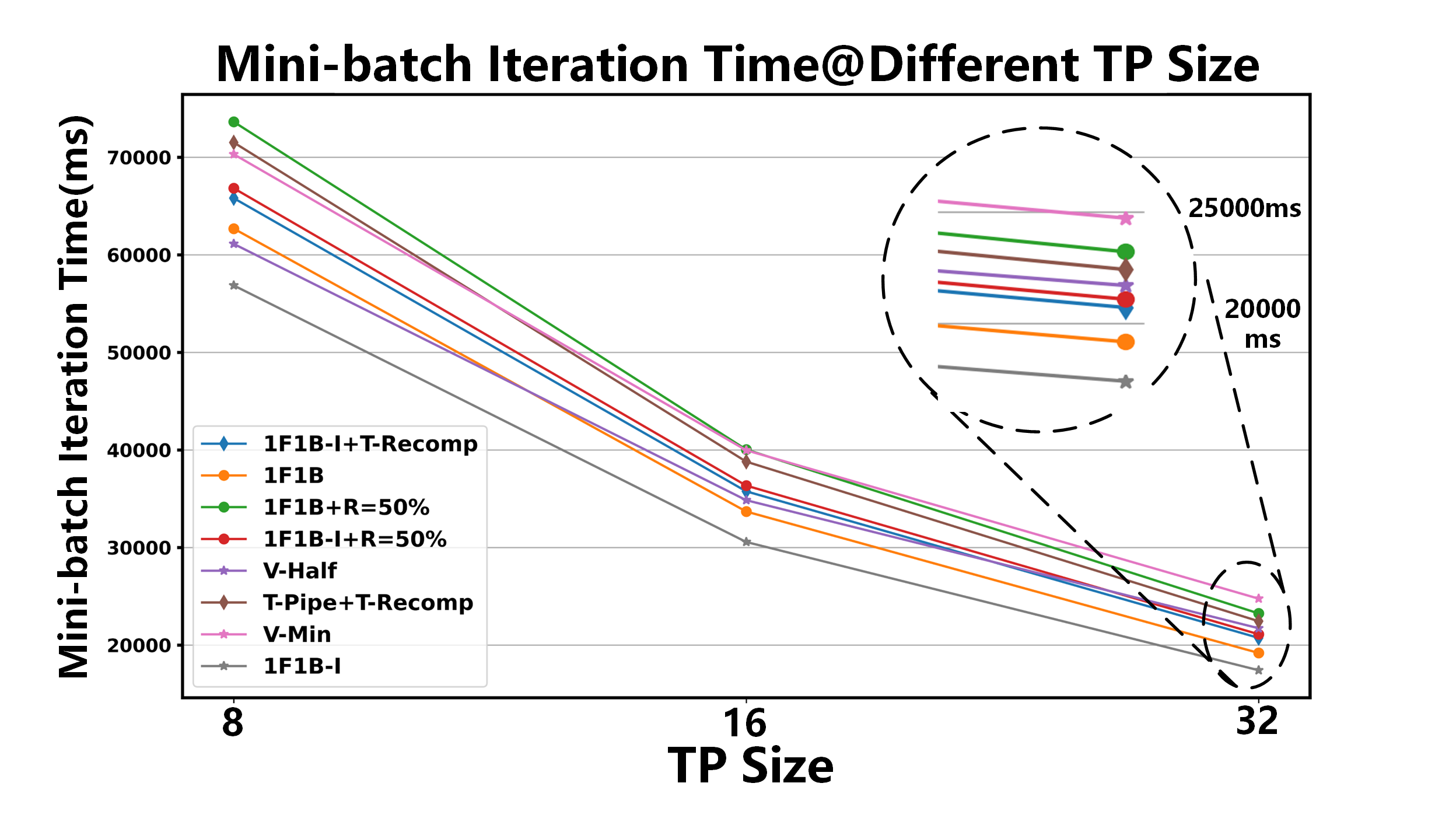}
    }
    \caption{Iteration time of different PP schedule on GPT3-175B when scaling TP size@PP8\_TP(8,16,32) on A100 cluster. Global batch=32, micro batch=1, seq len=16K. assume TP groups all communicate with high bandwidth NVLink.}
    \label{fig:ScaleUp}
\end{figure}

In addition to sequence length scaling, model size scaling is equally critical, requiring pipeline parallelism (PP) scheduling to maintain compatibility with large-scale tensor parallelism (TP) for dense models, as demonstrated in Fig. ~\ref{fig:ScaleUp}. As TP size increases, the TPipe solution exhibits scaling capabilities comparable to 1F1B and Interleave-1F1B(1F1B-I), while V-Half and V-Min show significant performance degradation. Compared to Interleave-1F1B(1F1B-I)+T-Recomp and T-Pipe+T-Recomp, V-Half and V-Min transition from a 7.67\%/1.72\% relative throughput advantage at TP8 to a 4.78\%/10.31\% deficit at TP32, primarily due to increased communication overhead. The backpropagation (BP) partitioning into BPA and BPW not only prevents overlapping of BPA's reduce-scatter operations but also requires additional all-gather operations for both activation gradients and activations in BPW. These compounded communication costs fundamentally limit their scalability along the TP dimension.

\subsection{Evaluation on Finer-grained Variants of TPipe}\label{DSE_ecaluation}

\begin{figure}[htbp]
    \centerline{
    \includegraphics[width=0.48\textwidth]{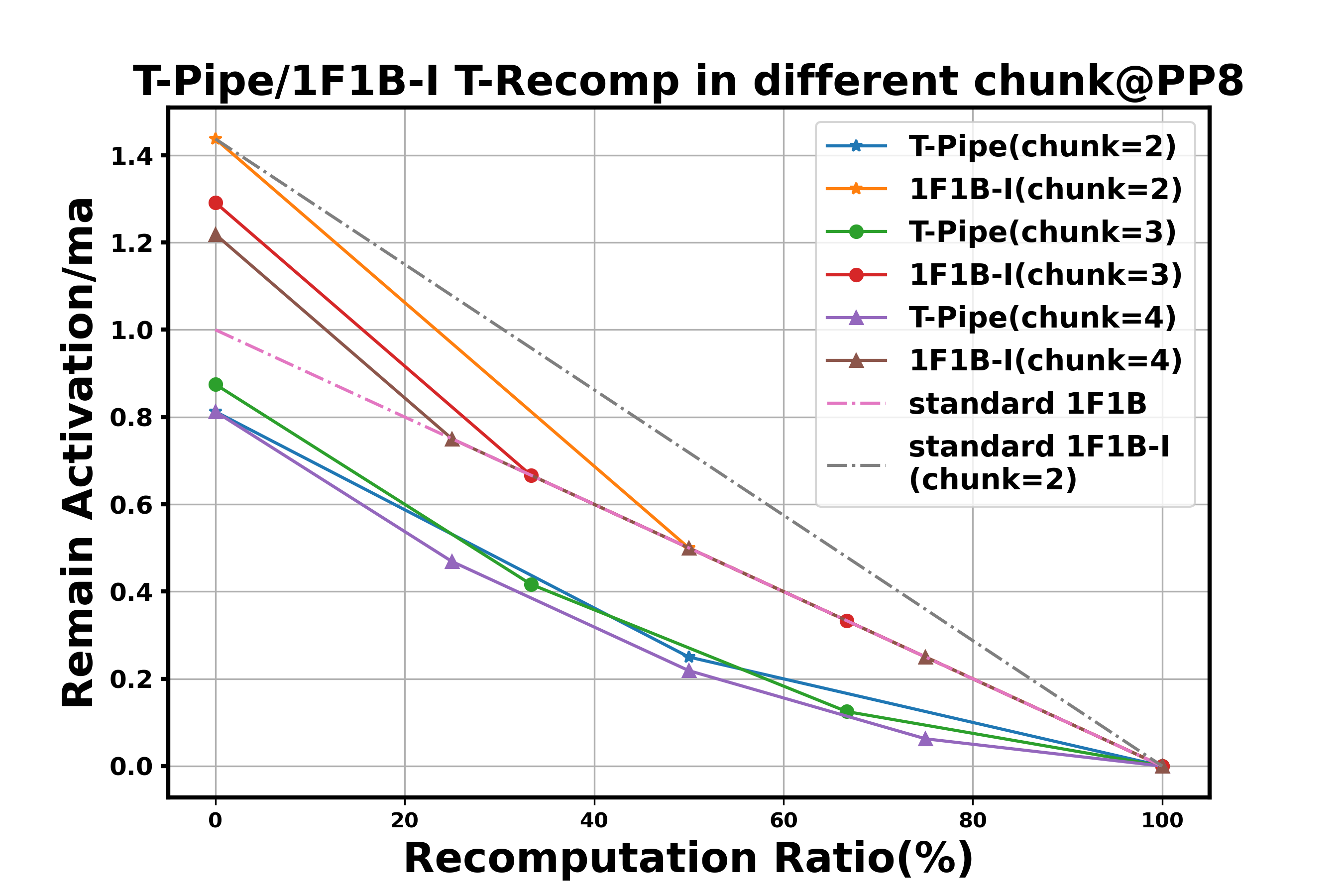}
    }
    \caption{in-depth analysis of T-Pipe/1F1B-I + T-Recomp. Additional buffer required for recomputation is ignored for better illustration.}
    \label{fig:E56}
\end{figure}

Fig. ~\ref{fig:E56} shows the effectiveness of T-Recomp across different chunk configurations in T-Pipe and Interleave-1F1B(1F1B-I). T-Pipe's locality-aware scheduling inherently achieves lower memory usage than 1F1B even without recomputation. When combined with T-Recomp, it further reduces activation storage at all recomputation ratios compared to standard 1F1B recomputation, with finer-grained chunks showing superior optimization due to more precise temporal locality application. While standard Interleave-1F1B underperforms 1F1B at all recomputation ratios due to significant extrinsic temporal overhead, T-Recomp fundamentally solves this problem: by discarding chunks with the poorest temporal locality, T-Recomp enables Interleave-1F1B to match standard 1F1B's recomputation efficiency, significantly improving its memory utilization.

\begin{figure}[htbp]
    \centerline{
    \includegraphics[width=0.48\textwidth]{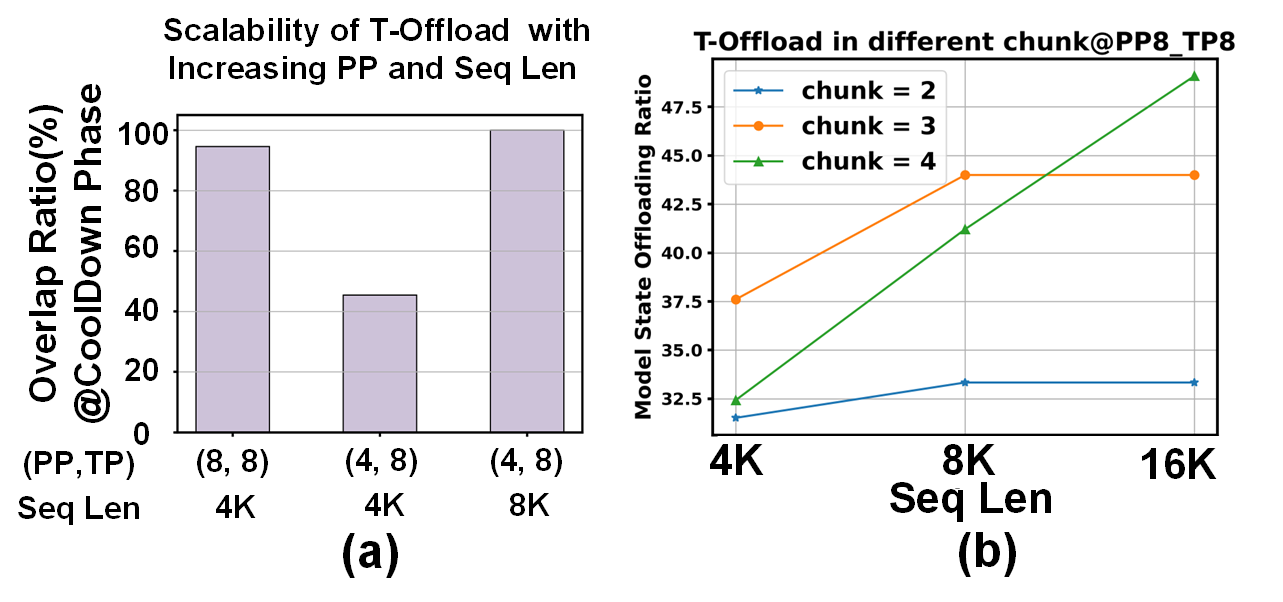}
    }
    \caption{(a)evaluate the scalability of T-Offload on a 16-layer model at
global batch=128, micro batch=2. (b)in-depth analysis of T-Offload}
    \label{fig:offload_DSE}
\end{figure}

We further verify the scalability of T-Offload, as illustrated in
Fig. ~\ref{fig:offload_DSE}(a). When training a 16-layer model with a sequence length of 4K on PP4\_TP8 setup, T-Pipe achieves only partial overlap (45.45\%) between the bubble created during the cooldown phase and the time required for T-Offload. However, when either the pipeline stage or the sequence length is doubled, the overlap improves significantly to 94.55\% and 100\%, respectively. Furthermore, T-Offload enables only half of the optimizers to be updated on GPU, leading to a slight decrease in overall training time. These results highlight the scalability of T-Offload, demonstrating its increasing effectiveness as the PP stage and sequence length grows.

Under multi-chunk partitioning, T-Pipe+T-Offload benefits from increased available bubbles. Fig. ~\ref{fig:offload_DSE}(b) demonstrates this progressive optimization: when sequence length grows from 4K to 8K, chunk 2 shows only marginal improvement in T-Offload efficiency since existing bubbles already suffice for its model state transfer. In contrast, chunks 4 exhibit significant gains as sequence length increases to 8K and beyond, since these chunks initially lack sufficient bubbles and require longer sequences to create adequate offloading windows. Chunk 3 exhibits intermediate behavior:at 8K sequences it reaches optimal efficiency with only the first chunk's poorest-temporal-locality model states remaining unoffloaded. When sequence length expands to 16K, chunks 2 and 3 already reach their optimization limits, reducing model state storage by 33.33\% and 44\% respectively, while chunk 4 achieves near-maximal 49.09\% reduction.

\section{RELATED WORK}

% \textbf{Hybrid Parallelism}. To address the challenges of scaling model size and sequence length, modern distributed training systems leverage GPU clusters to expand overall memory capacity while deploying various schedules. These strategies include data parallelism(DP), pipeline parallelism(PP), tensor parallelism(TP), context parallelism(CP), and so on. DP~\cite{ZeRO,FSDP}, PP~\cite{GPipe,DAPPLE,Interleave-1F1B,Hanayo,ZeroBubble}, and TP~\cite{AccPar,2D-TP,2p5D-TP,3D-TP} enable the distribution of model states across machines. CP~\cite{RingAttention,Ulysses} and sequence parallelism(SP)~\cite{SeqParallel} specialize in distributing activations across their respective dimensions. Hybrid parallelism combines these approaches, enabling efficient scaling to models with trillions of parameters. PP is an essential component of hybrid parallelism, particularly in cross-node deployments, due to its low communication requirement. However, PP suffers from bubbles and high activation memory overhead.

\textbf{Hybrid Parallelism}. Modern distributed training combines data (DP)~\cite{ZeRO,FSDP}, pipeline (PP)~\cite{GPipe,DAPPLE,Interleave-1F1B,Hanayo,ZeroBubble}, tensor (TP)~\cite{AccPar,2D-TP,2p5D-TP,3D-TP}, and context parallelism (CP)~\cite{RingAttention,Ulysses} to scale model size and sequence length. DP/PP/TP distribute model states, while CP/SP~\cite{SeqParallel} handle activations. Hybrid parallelism combines these approaches, enabling efficient scaling to models with trillions of parameters. PP’s low communication makes it critical for cross-node deployments. However, it suffers from bubbles and high activation memory overhead.

% \textbf{Pipeline Parallelism}. Most PP research focuses on reducing pipeline bubbles by finer-grained task partitioning ~\cite{GPipe, DAPPLE,Interleave-1F1B, Hanayo,ZeroBubble} or multi-dimensional task scheduling ~\cite{Chimera, Mixpipe}. However, they neglect the significant overhead of activation memory. ~\cite{CtrlMemPP} reduces activation lifespans through scheduling but risks introducing more bubbles and prolonging gradient lifespan. It's V-shaped scheduling also complicates compatibility with ZeRO-2/3 and T-Offload, making storage optimization for other data more challenging. ~\cite{AdaPipe} addresses activation memory imbalance with asymmetric partitioning but increases programming complexity. ~\cite{BreadthFirstPP,ZeroPP}, achieve ZeRO-2/3-compatibile PP at the cost of higher activation memory. ~\cite{LLAMA3} takes advantage of ~\cite{BreadthFirstPP} and 1F1B for improved ZeRO-2 compatibility but only maintains, rather than reduces, activation memory usage.

\textbf{Pipeline Parallelism}. Most PP research focuses on reducing pipeline bubbles by finer-grained task partitioning ~\cite{GPipe, DAPPLE,Interleave-1F1B, Hanayo,ZeroBubble} or multi-dimensional task scheduling ~\cite{Chimera, Mixpipe}. However, they neglect the significant overhead of activation memory. ~\cite{CtrlMemPP} reduces activation lifespans through scheduling but risks introducing more bubbles and prolonging gradient lifespan. ~\cite{AdaPipe} addresses activation memory imbalance with asymmetric partitioning but increases programming complexity. ~\cite{BreadthFirstPP,ZeroPP}, achieve ZeRO-2/3-compatibile PP at the cost of higher activation memory. ~\cite{LLAMA3} takes advantage of ~\cite{BreadthFirstPP} and 1F1B for improved ZeRO-2 compatibility but only maintains, rather than reduces, activation memory usage.

\textbf{Recomputation and Offloading}. Memory requirements can also be reduced by trading memory for computation or communication. For recomputation, some studies ~\cite{SeqParallel,MegatronKwai,SuperNeuron,CapuChin} focus on selective recomputation of different operators but fail to address the high memory usage of the projection operator. \textcolor{black}{~\cite{DTR} introduced temporal locality-aware recomputation principle and demonstrated its efficacy in DP, but this principle is still poorly explored in PP.} Some approaches leverage the characteristics of hybrid parallelism. ~\cite{AdaPipe} adapts recomputation to mitigate activation memory imbalance in PP. ~\cite{OverlapRecomp} overlaps recomputation with communication to reduce overhead. However, these approaches increase programming complexity. For offloading, ~\cite{Bpipe} addresses memory imbalance in PP by offloading activation across GPUs. Most studies focus on fully utilizing storage systems. For example, ~\cite{ZeRO-Offload,PatrickStar,Mobius} explore offloading activation and model states to DRAM, while ~\cite{ZeRO-Infinity,Angel-ptm,Smart-Infinity,Fuyou} extend this to NVMe. Due to limited bandwidth in offloading, these efforts primarily focus on identifying opportunities to overlap offloading with computation and realize near-data processing.

\section{CONCLUSION}

This work introduces TPipe to address the memory capacity limitations in LLM training. The core idea behind TPipe is to treat HBM as a fast but limited "cache," optimizing and leveraging temporal locality in LLM training for efficient HBM usage. T-Pipe enhances activation temporal locality in PP scheduling and, together with T-Recomp and T-Offload, discards data with poor temporal locality from HBM. Experiments show that T-Pipe can expand the trainable model size by 2.4x while maintaining comparable throughput, achieving 1.5x better than the 1F1B strategy combined with recomputation. At the same time, TPipe has good enough compatibility to make it suitable for large-scale hybrid parallelism training.


% \begin{acks}
% \end{acks}

\bibliographystyle{ACM-Reference-Format}
\bibliography{sample-base}

\begin{thebibliography}{10}
\providecommand{\url}[1]{#1}
\csname url@samestyle\endcsname
\providecommand{\newblock}{\relax}
\providecommand{\bibinfo}[2]{#2}
\providecommand{\BIBentrySTDinterwordspacing}{\spaceskip=0pt\relax}
\providecommand{\BIBentryALTinterwordstretchfactor}{4}
\providecommand{\BIBentryALTinterwordspacing}{\spaceskip=\fontdimen2\font plus
\BIBentryALTinterwordstretchfactor\fontdimen3\font minus \fontdimen4\font\relax}
\providecommand{\BIBforeignlanguage}[2]{{%
\expandafter\ifx\csname l@#1\endcsname\relax
\typeout{** WARNING: IEEEtran.bst: No hyphenation pattern has been}%
\typeout{** loaded for the language `#1'. Using the pattern for}%
\typeout{** the default language instead.}%
\else
\language=\csname l@#1\endcsname
\fi
#2}}
\providecommand{\BIBdecl}{\relax}
\BIBdecl

\bibitem{ScalingLaw}
J.~Kaplan, S.~McCandlish, T.~Henighan, T.~B. Brown, B.~Chess, R.~Child, S.~Gray, A.~Radford, J.~Wu, and D.~Amodei, ``Scaling laws for neural language models,'' \emph{arXiv preprint arXiv:2001.08361}, 2020.

\bibitem{GPT1}
A.~Radford, K.~Narasimhan, T.~Salimans, and I.~Sutskever, ``Improving language understanding by generative pre-training,'' 2018.

\bibitem{LLAMA3}
A.~Dubey, A.~Jauhri, A.~Pandey, A.~Kadian, A.~Al-Dahle, A.~Letman, A.~Mathur, A.~Schelten, A.~Yang, A.~Fan \emph{et~al.}, ``The llama 3 herd of models,'' \emph{arXiv preprint arXiv:2407.21783}, 2024.

\bibitem{GPipe}
\BIBentryALTinterwordspacing
Y.~Huang, Y.~Cheng, A.~Bapna, O.~Firat, M.~X. Chen, D.~Chen, H.~Lee, J.~Ngiam, Q.~V. Le, Y.~Wu, and Z.~Chen, ``Gpipe: Efficient training of giant neural networks using pipeline parallelism,'' 2019. [Online]. Available: \url{https://arxiv.org/abs/1811.06965}
\BIBentrySTDinterwordspacing

\bibitem{DAPPLE}
\BIBentryALTinterwordspacing
S.~Fan, Y.~Rong, C.~Meng, Z.~Cao, S.~Wang, Z.~Zheng, C.~Wu, G.~Long, J.~Yang, L.~Xia, L.~Diao, X.~Liu, and W.~Lin, ``Dapple: a pipelined data parallel approach for training large models,'' in \emph{Proceedings of the 26th ACM SIGPLAN Symposium on Principles and Practice of Parallel Programming}, ser. PPoPP '21.\hskip 1em plus 0.5em minus 0.4em\relax New York, NY, USA: Association for Computing Machinery, 2021, p. 431–445. [Online]. Available: \url{https://doi.org/10.1145/3437801.3441593}
\BIBentrySTDinterwordspacing

\bibitem{Interleave-1F1B}
\BIBentryALTinterwordspacing
D.~Narayanan, M.~Shoeybi, J.~Casper, P.~LeGresley, M.~Patwary, V.~Korthikanti, D.~Vainbrand, P.~Kashinkunti, J.~Bernauer, B.~Catanzaro, A.~Phanishayee, and M.~Zaharia, ``Efficient large-scale language model training on gpu clusters using megatron-lm,'' in \emph{Proceedings of the International Conference for High Performance Computing, Networking, Storage and Analysis}, ser. SC '21.\hskip 1em plus 0.5em minus 0.4em\relax New York, NY, USA: Association for Computing Machinery, 2021. [Online]. Available: \url{https://doi.org/10.1145/3458817.3476209}
\BIBentrySTDinterwordspacing

\bibitem{Hanayo}
Z.~Liu, S.~Cheng, H.~Zhou, and Y.~You, ``Hanayo: Harnessing wave-like pipeline parallelism for enhanced large model training efficiency,'' in \emph{Proceedings of the International Conference for High Performance Computing, Networking, Storage and Analysis}, 2023, pp. 1--13.

\bibitem{ZeroBubble}
P.~Qi, X.~Wan, G.~Huang, and M.~Lin, ``Zero bubble pipeline parallelism,'' \emph{arXiv preprint arXiv:2401.10241}, 2023.

\bibitem{Chimera}
S.~Li and T.~Hoefler, ``Chimera: efficiently training large-scale neural networks with bidirectional pipelines,'' in \emph{Proceedings of the International Conference for High Performance Computing, Networking, Storage and Analysis}, 2021, pp. 1--14.

\bibitem{Mixpipe}
W.~Zhang, B.~Zhou, X.~Tang, Z.~Wang, and S.~Hu, ``Mixpipe: Efficient bidirectional pipeline parallelism for training large-scale models,'' in \emph{2023 60th ACM/IEEE Design Automation Conference (DAC)}.\hskip 1em plus 0.5em minus 0.4em\relax IEEE, 2023, pp. 1--6.

\bibitem{MegatronKwai}
T.~Yuan, Y.~Liu, X.~Ye, S.~Zhang, J.~Tan, B.~Chen, C.~Song, and D.~Zhang, ``Accelerating the training of large language models using efficient activation rematerialization and optimal hybrid parallelism,'' in \emph{2024 USENIX Annual Technical Conference (USENIX ATC 24)}, 2024, pp. 545--561.

\bibitem{Bpipe}
T.~Kim, H.~Kim, G.-I. Yu, and B.-G. Chun, ``Bpipe: memory-balanced pipeline parallelism for training large language models,'' in \emph{International Conference on Machine Learning}.\hskip 1em plus 0.5em minus 0.4em\relax PMLR, 2023, pp. 16\,639--16\,653.

\bibitem{ZeRO}
S.~Rajbhandari, J.~Rasley, O.~Ruwase, and Y.~He, ``Zero: Memory optimizations toward training trillion parameter models,'' in \emph{SC20: International Conference for High Performance Computing, Networking, Storage and Analysis}.\hskip 1em plus 0.5em minus 0.4em\relax IEEE, 2020, pp. 1--16.

\bibitem{BreadthFirstPP}
J.~Lamy-Poirier, ``Breadth-first pipeline parallelism,'' \emph{Proceedings of Machine Learning and Systems}, vol.~5, pp. 48--67, 2023.

\bibitem{ZeroPP}
\BIBentryALTinterwordspacing
D.~Tang, L.~Jiang, J.~Zhou, M.~Jin, H.~Li, X.~Zhang, Z.~Pei, and J.~Zhai, ``Zeropp: Unleashing exceptional parallelism efficiency through tensor-parallelism-free methodology,'' 2024. [Online]. Available: \url{https://arxiv.org/abs/2402.03791}
\BIBentrySTDinterwordspacing

\bibitem{ChainTrainingMem}
T.~Chen, B.~Xu, C.~Zhang, and C.~Guestrin, ``Training deep nets with sublinear memory cost,'' \emph{arXiv preprint arXiv:1604.06174}, 2016.

\bibitem{SeqParallel}
V.~A. Korthikanti, J.~Casper, S.~Lym, L.~McAfee, M.~Andersch, M.~Shoeybi, and B.~Catanzaro, ``Reducing activation recomputation in large transformer models,'' \emph{Proceedings of Machine Learning and Systems}, vol.~5, pp. 341--353, 2023.

\bibitem{ZeRO-Offload}
\BIBentryALTinterwordspacing
J.~Ren, S.~Rajbhandari, R.~Y. Aminabadi, O.~Ruwase, S.~Yang, M.~Zhang, D.~Li, and Y.~He, ``Zero-offload: Democratizing billion-scale model training,'' 2021. [Online]. Available: \url{https://arxiv.org/abs/2101.06840}
\BIBentrySTDinterwordspacing

\bibitem{FSDP}
\BIBentryALTinterwordspacing
Y.~Zhao, A.~Gu, R.~Varma, L.~Luo, C.-C. Huang, M.~Xu, L.~Wright, H.~Shojanazeri, M.~Ott, S.~Shleifer, A.~Desmaison, C.~Balioglu, P.~Damania, B.~Nguyen, G.~Chauhan, Y.~Hao, A.~Mathews, and S.~Li, ``Pytorch fsdp: Experiences on scaling fully sharded data parallel,'' 2023. [Online]. Available: \url{https://arxiv.org/abs/2304.11277}
\BIBentrySTDinterwordspacing

\bibitem{AccPar}
L.~Song, F.~Chen, Y.~Zhuo, X.~Qian, H.~Li, and Y.~Chen, ``Accpar: Tensor partitioning for heterogeneous deep learning accelerators,'' in \emph{2020 IEEE International Symposium on High Performance Computer Architecture (HPCA)}.\hskip 1em plus 0.5em minus 0.4em\relax IEEE, 2020, pp. 342--355.

\bibitem{2D-TP}
\BIBentryALTinterwordspacing
Q.~Xu, S.~Li, C.~Gong, and Y.~You, ``An efficient 2d method for training super-large deep learning models,'' 2021. [Online]. Available: \url{https://arxiv.org/abs/2104.05343}
\BIBentrySTDinterwordspacing

\bibitem{2p5D-TP}
B.~Wang, Q.~Xu, Z.~Bian, and Y.~You, ``Tesseract: Parallelize the tensor parallelism efficiently,'' in \emph{Proceedings of the 51st International Conference on Parallel Processing}, 2022, pp. 1--11.

\bibitem{3D-TP}
Z.~Bian, Q.~Xu, B.~Wang, and Y.~You, ``Maximizing parallelism in distributed training for huge neural networks,'' \emph{arXiv preprint arXiv:2105.14450}, 2021.

\bibitem{RingAttention}
H.~Liu, M.~Zaharia, and P.~Abbeel, ``Ring attention with blockwise transformers for near-infinite context,'' \emph{arXiv preprint arXiv:2310.01889}, 2023.

\bibitem{Ulysses}
S.~A. Jacobs, M.~Tanaka, C.~Zhang, M.~Zhang, S.~L. Song, S.~Rajbhandari, and Y.~He, ``Deepspeed ulysses: System optimizations for enabling training of extreme long sequence transformer models,'' \emph{arXiv preprint arXiv:2309.14509}, 2023.

\bibitem{CtrlMemPP}
P.~Qi, X.~Wan, N.~Amar, and M.~Lin, ``Pipeline parallelism with controllable memory,'' \emph{arXiv preprint arXiv:2405.15362}, 2024.

\bibitem{AdaPipe}
Z.~Sun, H.~Cao, Y.~Wang, G.~Feng, S.~Chen, H.~Wang, and W.~Chen, ``Adapipe: Optimizing pipeline parallelism with adaptive recomputation and partitioning,'' in \emph{Proceedings of the 29th ACM International Conference on Architectural Support for Programming Languages and Operating Systems, Volume 3}, 2024, pp. 86--100.

\bibitem{OverlapRecomp}
\BIBentryALTinterwordspacing
P.~Chen, W.~Zhang, S.~He, Y.~Gu, Z.~Peng, K.~Huang, X.~Zhan, W.~Chen, Y.~Zheng, Z.~Wang, Y.~Yin, and G.~Chen, ``Optimizing large model training through overlapped activation recomputation,'' 2024. [Online]. Available: \url{https://arxiv.org/abs/2406.08756}
\BIBentrySTDinterwordspacing

\bibitem{PatrickStar}
J.~Fang, Z.~Zhu, S.~Li, H.~Su, Y.~Yu, J.~Zhou, and Y.~You, ``Parallel training of pre-trained models via chunk-based dynamic memory management,'' \emph{IEEE Transactions on Parallel and Distributed Systems}, vol.~34, no.~1, pp. 304--315, 2022.

\bibitem{Mobius}
Y.~Feng, M.~Xie, Z.~Tian, S.~Wang, Y.~Lu, and J.~Shu, ``Mobius: Fine tuning large-scale models on commodity gpu servers,'' in \emph{Proceedings of the 28th ACM International Conference on Architectural Support for Programming Languages and Operating Systems, Volume 2}, 2023, pp. 489--501.

\bibitem{ZeRO-Infinity}
S.~Rajbhandari, O.~Ruwase, J.~Rasley, S.~Smith, and Y.~He, ``Zero-infinity: Breaking the gpu memory wall for extreme scale deep learning,'' in \emph{Proceedings of the international conference for high performance computing, networking, storage and analysis}, 2021, pp. 1--14.

\bibitem{Angel-ptm}
X.~Nie, Y.~Liu, F.~Fu, J.~Xue, D.~Jiao, X.~Miao, Y.~Tao, and B.~Cui, ``Angel-ptm: A scalable and economical large-scale pre-training system in tencent,'' \emph{arXiv preprint arXiv:2303.02868}, 2023.

\bibitem{Smart-Infinity}
H.~Jang, J.~Song, J.~Jung, J.~Park, Y.~Kim, and J.~Lee, ``Smart-infinity: Fast large language model training using near-storage processing on a real system,'' in \emph{2024 IEEE International Symposium on High-Performance Computer Architecture (HPCA)}.\hskip 1em plus 0.5em minus 0.4em\relax IEEE, 2024, pp. 345--360.

\bibitem{Fuyou}
C.~Liao, M.~Sun, Z.~Yang, K.~Chen, B.~Yuan, F.~Wu, and Z.~Wang, ``Adding nvme ssds to enable and accelerate 100b model fine-tuning on a single gpu,'' \emph{arXiv preprint arXiv:2403.06504}, 2024.

\bibitem{LLM}
\BIBentryALTinterwordspacing
T.~B. Brown, B.~Mann, N.~Ryder, M.~Subbiah, J.~Kaplan, P.~Dhariwal, A.~Neelakantan, P.~Shyam, G.~Sastry, A.~Askell, S.~Agarwal, A.~Herbert-Voss, G.~Krueger, T.~Henighan, R.~Child, A.~Ramesh, D.~M. Ziegler, J.~Wu, C.~Winter, C.~Hesse, M.~Chen, E.~Sigler, M.~Litwin, S.~Gray, B.~Chess, J.~Clark, C.~Berner, S.~McCandlish, A.~Radford, I.~Sutskever, and D.~Amodei, ``Language models are few-shot learners,'' 2020. [Online]. Available: \url{https://arxiv.org/abs/2005.14165}
\BIBentrySTDinterwordspacing

\bibitem{Multimodal}
\BIBentryALTinterwordspacing
A.~Radford, J.~W. Kim, C.~Hallacy, A.~Ramesh, G.~Goh, S.~Agarwal, G.~Sastry, A.~Askell, P.~Mishkin, J.~Clark, G.~Krueger, and I.~Sutskever, ``Learning transferable visual models from natural language supervision,'' in \emph{Proceedings of the 38th International Conference on Machine Learning}, ser. Proceedings of Machine Learning Research, M.~Meila and T.~Zhang, Eds., vol. 139.\hskip 1em plus 0.5em minus 0.4em\relax PMLR, 18--24 Jul 2021, pp. 8748--8763. [Online]. Available: \url{https://proceedings.mlr.press/v139/radford21a.html}
\BIBentrySTDinterwordspacing

\bibitem{microbump}
S.~W. Yoon, J.~H. Ku, N.~Suthiwongsunthorn, P.~C. Marimuthu, and F.~Carson, ``Fabrication and packaging of microbump interconnections for 3d tsv,'' in \emph{2009 IEEE International Conference on 3D System Integration}.\hskip 1em plus 0.5em minus 0.4em\relax IEEE, 2009, pp. 1--5.

\bibitem{TSV}
N.~Khan, V.~S. Rao, S.~Lim, H.~S. We, V.~Lee, X.~Zhang, E.~B. Liao, R.~Nagarajan, T.~C. Chai, V.~Kripesh, and J.~H. Lau, ``Development of 3-d silicon module with tsv for system in packaging,'' \emph{IEEE Transactions on Components and Packaging Technologies}, vol.~33, no.~1, pp. 3--9, 2010.

\bibitem{HybridBonding}
H.~Lee, J.~Kim, M.-K. Kim, W.~Lee, A.~Jang, H.~Lee, and D.-W. Kim, ``A study on d2w hybrid cu bonding technology for hbm multi-die stacking,'' in \emph{2024 IEEE 74th Electronic Components and Technology Conference (ECTC)}.\hskip 1em plus 0.5em minus 0.4em\relax IEEE, 2024, pp. 76--80.

\bibitem{HybridBonding_Problem}
K.~Kim, S.~Lim, D.~Jung, J.~Choi, S.~Na, J.~Yeom, M.~Lee, J.~Kim, J.~Kwon, K.-I. Moon, G.~Lee, and K.~Lee, ``C2w hybrid bonding interconnect technology for higher density and better thermal dissipation of high bandwidth memory,'' in \emph{2023 IEEE 73rd Electronic Components and Technology Conference (ECTC)}.\hskip 1em plus 0.5em minus 0.4em\relax IEEE, 2023, pp. 1048--1052.

\bibitem{DeepSpeed}
J.~Rasley, S.~Rajbhandari, O.~Ruwase, and Y.~He, ``Deepspeed: System optimizations enable training deep learning models with over 100 billion parameters,'' in \emph{Proceedings of the 26th ACM SIGKDD International Conference on Knowledge Discovery \& Data Mining}, 2020, pp. 3505--3506.

\bibitem{H100}
\BIBentryALTinterwordspacing
``Nvidia h100 tensor core gpu architecture,'' 3 2022, [Online]. Available: \url{https://resources.nvidia.com/en-us-tensor-core/gtc22-whitepaper-hopper}. [Online]. Available: \url{https://resources.nvidia.com/en-us-tensor-core/gtc22-whitepaper-hopper}
\BIBentrySTDinterwordspacing

\bibitem{PipeFisher}
K.~Osawa, S.~Li, and T.~Hoefler, ``Pipefisher: Efficient training of large language models using pipelining and fisher information matrices,'' \emph{Proceedings of Machine Learning and Systems}, vol.~5, pp. 708--727, 2023.

\bibitem{Flashattention}
T.~Dao, D.~Fu, S.~Ermon, A.~Rudra, and C.~R{\'e}, ``Flashattention: Fast and memory-efficient exact attention with io-awareness,'' \emph{Advances in neural information processing systems}, vol.~35, pp. 16\,344--16\,359, 2022.

\bibitem{LLAMA2}
H.~Touvron, L.~Martin, K.~Stone, P.~Albert, A.~Almahairi, Y.~Babaei, N.~Bashlykov, S.~Batra, P.~Bhargava, S.~Bhosale \emph{et~al.}, ``Llama 2: Open foundation and fine-tuned chat models,'' \emph{arXiv preprint arXiv:2307.09288}, 2023.

\bibitem{OPT}
S.~Zhang, S.~Roller, N.~Goyal, M.~Artetxe, M.~Chen, S.~Chen, C.~Dewan, M.~Diab, X.~Li, X.~V. Lin \emph{et~al.}, ``Opt: Open pre-trained transformer language models,'' \emph{arXiv preprint arXiv:2205.01068}, 2022.

\bibitem{PaLM}
A.~Chowdhery, S.~Narang, J.~Devlin, M.~Bosma, G.~Mishra, A.~Roberts, P.~Barham, H.~W. Chung, C.~Sutton, S.~Gehrmann \emph{et~al.}, ``Palm: Scaling language modeling with pathways,'' \emph{Journal of Machine Learning Research}, vol.~24, no. 240, pp. 1--113, 2023.

\bibitem{Qwen}
A.~Yang, B.~Yang, B.~Zhang, B.~Hui, B.~Zheng, B.~Yu, C.~Li, D.~Liu, F.~Huang, H.~Wei \emph{et~al.}, ``Qwen2. 5 technical report,'' \emph{arXiv preprint arXiv:2412.15115}, 2024.

\bibitem{DTR}
M.~Kirisame, S.~Lyubomirsky, A.~Haan, J.~Brennan, M.~He, J.~Roesch, T.~Chen, and Z.~Tatlock, ``Dynamic tensor rematerialization,'' \emph{arXiv preprint arXiv:2006.09616}, 2020.

\bibitem{SuperNeuron}
L.~Wang, J.~Ye, Y.~Zhao, W.~Wu, A.~Li, S.~L. Song, Z.~Xu, and T.~Kraska, ``Superneurons: Dynamic gpu memory management for training deep neural networks,'' in \emph{Proceedings of the 23rd ACM SIGPLAN symposium on principles and practice of parallel programming}, 2018, pp. 41--53.

\bibitem{CapuChin}
X.~Peng, X.~Shi, H.~Dai, H.~Jin, W.~Ma, Q.~Xiong, F.~Yang, and X.~Qian, ``Capuchin: Tensor-based gpu memory management for deep learning,'' in \emph{Proceedings of the Twenty-Fifth International Conference on Architectural Support for Programming Languages and Operating Systems}, 2020, pp. 891--905.

\bibitem{SimAI}
X.~Wang, Q.~Li, Y.~Xu, G.~Lu, D.~Li, L.~Chen, H.~Zhou, L.~Zheng, S.~Zhang, Y.~Zhu \emph{et~al.}, ``$\{$SimAI$\}$: Unifying architecture design and performance tuning for $\{$Large-Scale$\}$ large language model training with scalability and precision,'' in \emph{22nd USENIX Symposium on Networked Systems Design and Implementation (NSDI 25)}, 2025, pp. 541--558.

\bibitem{Dojo}
B.~Chang, R.~Kurian, D.~Williams, and E.~Quinnell, ``Dojo: Super-compute system scaling for ml training,'' in \emph{2022 IEEE Hot Chips 34 Symposium (HCS)}.\hskip 1em plus 0.5em minus 0.4em\relax IEEE Computer Society, 2022, pp. 1--45.

\bibitem{Cerebras}
S.~Lie, ``Cerebras architecture deep dive: First look inside the hw/sw co-design for deep learning: Cerebras systems,'' in \emph{2022 IEEE Hot Chips 34 Symposium (HCS)}.\hskip 1em plus 0.5em minus 0.4em\relax IEEE Computer Society, 2022, pp. 1--34.

\bibitem{GPT3}
T.~Brown, B.~Mann, N.~Ryder, M.~Subbiah, J.~D. Kaplan, P.~Dhariwal, A.~Neelakantan, P.~Shyam, G.~Sastry, A.~Askell \emph{et~al.}, ``Language models are few-shot learners,'' \emph{Advances in neural information processing systems}, vol.~33, pp. 1877--1901, 2020.

\end{thebibliography}

\appendix

\section{\textbf{Theoretical Analysis of T-Pipe}}\label{Appendix T-Pipe} 

\begin{figure}[htbp]
    \centerline{
    \includegraphics[width=0.48\textwidth]{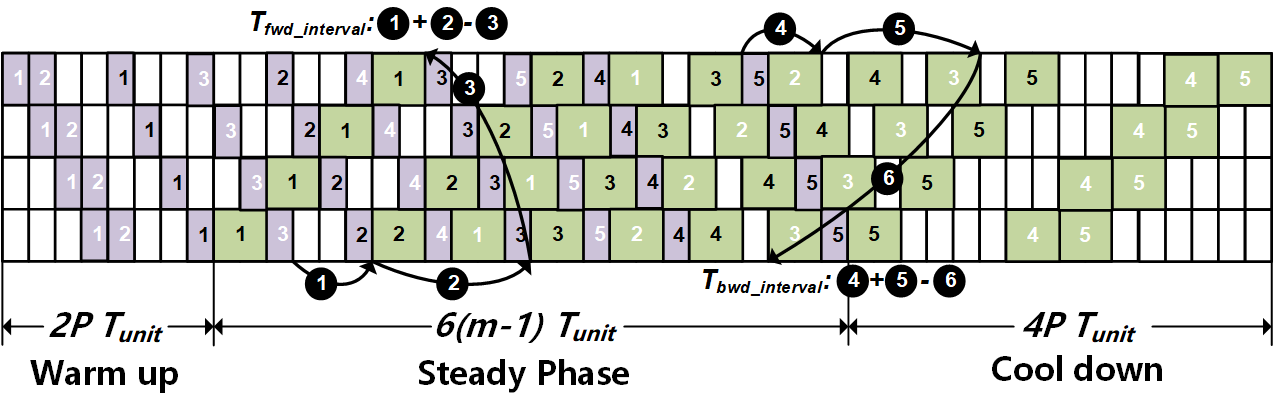}
    }
    \caption{Theoretical Analysis of T-Pipe.}
    \label{fig:T-Pipe_Analysis}
\end{figure}

A further theoretical analysis of T-Pipe on activation storage and bubble ratio is provided as follow. For simplicity, we assume $T_{bwd}=2T_{fwd}$, basic unit of time $T_{unit}$  in the pipeline scheduling diagram (e.g., Fig. ~\ref{fig:T-Pipe_Analysis}) is $\frac{T_{fwd}}{2p}$, and the activation storage for each task is $\frac{m_a}{2p}$.

In T-Pipe, activation storage can be estimated by analyzing the activation lifespan, approximately 75\%  of $m_a$. As shown in Fig. ~\ref{fig:T-Pipe_Analysis}, Stage 0 launches chunk 2 of the fourth micro-batch before releasing the activation storage from chunk 2 of the third micro-batch. Since tasks on a single stage execute in cycles of $6T_{unit}$, the accumulated activation storage for chunk 2 can be represented as $\lceil \frac{T_{life\_chunk2}}{6T_{unit}} \rceil$, where $T_{life\_chunk2}$ denotes the lifespan of chunk 2's activations, precisely the time from the completion of its forward pass to the beginning of the corresponding backward pass at PP stage 0. Similarly, the accumulation of activation storage for chunk 1 can be expressed as $\lceil  
\frac{T_{life\_chunk1}}{6T_{unit}} \rceil$. Additionally, we observe that the forward execution times for both chunks $T_{fwd\_chunk1}$, $T_{fwd\_chunk2}$ are each $pT_{unit}$, while the backward execution times $T_{bwd\_chunk1}$, $T_{bwd\_chunk2}$ are each $2pT_{unit}$. There is also an extra interval $T_{fwd\_interval}$ between the forward chunks, derived from steps 1, 2, and 3 in Fig. ~\ref{fig:T-Pipe_Analysis} as $(3+6 \lceil \frac{(p-3)}{6} \rceil -p)T_{unit}$. Similarly, an additional interval $T_{bwd\_interval}$ exists between the backward chunks, derived from steps 4, 5, and 6 in Fig. ~\ref{fig:T-Pipe_Analysis} as $(3+6\lceil \frac{(2p-3)}{6} \rceil-2p)T_{unit}$. Thus, for Stage 0, the lifespans of the activations for chunk 1 and chunk 2, denoted $T_{life\_chunk1}$,$T_{life\_chunk2}$ can be expressed as:
\begin{equation}
T_{life\_chunk2}=T_{fwd\_chunk2}+T_{bwd\_chunk2}-2T_{unit}\label{eq}
\end{equation}

\begin{equation}
\begin{aligned}
T_{life\_chunk1}=&T_{fwd\_chunk1}+T_{fwd\_interval}+\\
&T_{life\_chunk2}+T_{bwd\_interval}+T_{bwd\_chunk1}\label{eq}
\end{aligned}
\end{equation}
As a result, the peak activation storage requirements for chunks 1 and 2 are $\lceil \frac{2}{3} +\lceil \frac{(p-3)}{6} \rceil +\lceil \frac{(2p-3)}{6} \rceil+ \frac{p}{2} \rceil \frac{m_a}{2p}$,$\lceil \frac{(3p-2)}{6}\rceil  \frac{m_a}{2p} $ respectively. When P is large, these values approach $~\frac{m_a}{2},\frac{m_a}{4}$, making the total peak activation storage approximately 75\% of $m_a$.

In addition, T-Pipe exhibits a bubble ratio similar to 1F1B. As depicted in Fig. ~\ref{fig:T-Pipe_Analysis}, its execution comprises three phases: $T_{warmup}, T_{steady}$ and $T_{cooldown}$.

$\bullet$ The warmup phase includes the forward pass of the first micro-batch, calculated as $2p T_{unit}$.

$\bullet$ The Steady phase covers the forward pass of chunk 1 for the remaining $(m - 1)$ microbatches, totaling $6(m - 1)T_{unit}$.

$\bullet$ The cooldown phase mainly involves the backward pass of the last micro-batch, calculated as $4p T_{unit}$.

The total execution time aggregates three phases into $6(m+p-1)T_{unit}$, with bubble time accounting for $6(p-1)T_{unit}$. Consequently, the bubble ratio is $\frac{p-1}{m+p-1}$—identical to 1F1B when P2P communication costs is ignored. In fact, the partition of workload into two chunks requires twice the P2P communication rounds compared to 1F1B. Due to increased P2P communication in T-Pipe, the Model Flop Utilization is slightly reduced.

\section{Formalization on Delay Rounds at T-Recomp}\label{Appendix A}

In T-Pipe scheduling, an additional interval time $T_{fwd\_interval}$ exists between forward chunks, derived from steps 1, 2, and 3 in Fig.~\ref{fig:T-Pipe_Analysis}, resulting in $(3+6\lceil \frac{(p-3)}{6} \rceil-p)T_{unit}$. In T-Recomp’s shallow-layer full recomputation, the execution times for steps 1, 2, and 3 respectively change $T_{unit},\lceil \frac{(p-3)}{6} \rceil T_{unit}$, and $\lceil \frac{(p-1)}{2} \rceil T_{unit}$. Therefore, $\Delta T_{fwd\_interval}$ becomes $(\lceil \frac{(p-1)}{2} \rceil-\lceil \frac{(p-3)}{6} \rceil-1) T_{unit}(P\geq3)$. Based on this, We can formalize the number of rounds k required to delay the launch of chunk 2 as an optimization problem for $P\geq3$:

\begin{equation}
\begin{aligned}
Min & \quad  k \\
s.t. \quad T_{fwd\_interval} - & \Delta T_{fwd\_interval} + 7kT_{unit} \geq 0 \\
k & \in N \\
p & \geq 3
\end{aligned}
\end{equation}
Therefore, when $8 \leq p \leq 40$, we have $k=1$, meaning that chunk 2 needs to be delayed by one round.

\section{\textbf{Theoretical Analysis of T-Recomp}}\label{Appendix T-Recomp}
We analyze activation storage requirements using full recomputation of shallow layers as an example. With a chunk size of 2, the remaining activation storage is capped at only 25\% of $m_a$. Observing micro-batch 2 in Fig. ~\ref{fig:6}(d), the total intervals introduced during the forward execution of chunk 2 is $\lceil \frac{(p-1)}{2} \rceil T_{unit}$. Therefore, for Stage 0, the lifespan of chunk 2's activations is given by:
\begin{equation}
T_{life\_chunk2} = (3p+\lceil \frac{(p-1)}{2} \rceil-2) T_{unit}\label{eq}
\end{equation}

Here, $(3p-2) T_{unit}$ represents the intrinsic temporal overhead of chunk 2. Moreover, since tasks in pipeline parallelism repeat with a period of $7T_{unit}$, the required number of activation storage blocks for chunk 2 is $\lceil \frac{T_{life\_chunk2}}{7T_{unit}} \rceil$, which simplifies to $\lfloor \frac{p}{2} \rfloor$. Taking into account that the recomputation buffer requires an additional block of space for regenerated activation values, we need $\lfloor \frac{p}{2} \rfloor + 1$ blocks for activation. Thus, with T-Recomp's full recomputation for chunk 1, the activation storage is only 25\% of $m_a$. By comparison, standard recomputation in 1F1B with the same computation budget requires 50\% of $m_a$ for storage, making T-Recomp 1.5 times more storage-efficient than standard recomputation.

Execution time of T-Pipe+T-Recomp is similar to 1F1B with recomputation, as shown in Fig. ~\ref{fig:6}(d). The warm-up phase/cool-down phase take $2pT_{unit}$/$4pT_{unit}$ respectively. The steady phase involves the forward computation of chunk 2 for the remaining $(m-1)$ microbatches, consuming $7(m-1)T_{unit}$. This results in a total execution time of $[6p+7(m-1)] T_{unit}$. By comparison, the execution time for 1F1B with 50\% recomputation is $7(m-1+p)T_{unit}$. Notably, T-Recomp achieves 50\% P2P communication overlap during forward computation, resulting in similar P2P communication impacts between T-Pipe+T-Recomp and 1F1B.

\section{\textbf{Scalability of T-Offload with Increasing DP, PP, and Sequence Length}}\label{Appendix T-Offload}

Two conditions must be met to complete the T-Offload process within the available 'bubbles.' These conditions are more easily satisfied as the degree of PP and sequence length increase. The first condition requires the gradient offload and the CPU optimizer updates to be completed within the cooldown phase's 'bubbles.' In T-Pipe, the interval between backward computations of chunks is $(3+6\lceil \frac{(2p-3)}{6} \rceil-2p)T_{unit}$, which results in a delay of $\lceil \frac{(2p-3)}{6} \rceil$ rounds for chunk 1 relative to chunk 2. This aligns the backward computation of the first micro-batch for chunk 1 with the (1+$\lceil \frac{(2p-3)}{6} \rceil$) th one of chunk 2. Thus, we define:

\begin{equation}
T_{avaliable\_offload}= (p-\lceil \frac{(2p-3)}{6} \rceil-1)\frac{T_{bwd}}{2p}\label{eq}
\end{equation}

Here, $T_{avaliable\_offload}$ represents the idle time available between the end of the backward computation for chunk 2 and the start of backward computation for chunk 1 during the cooldown phase. Ideally, this time would be sufficient to complete the gradient offload and the CPU update calculations. It follows that:

\begin{equation}
\frac {T_{step}}{2p} \leq T_{avaliable\_offload}  \Rightarrow  \frac {T_{step}}{T_{bwd}} \frac {1}{(p-\lceil \frac{(2p-3)}{6} \rceil-1)} \leq 1\label{eq}
\end{equation}

Where $T_{step}$ denotes the total time required for offloading gradients in all layers of DNN and performing CPU optimizer updates. As $P$ increases, the impact of the constant terms decreases, simplifying the equation to $\frac{T_{step}}{T_{bwd}}   \frac{3}{2p} \leq 1$ for large $P$, nearly offering a linear increase in idle time, thus facilitating easier fulfillment of this condition. The ratio $\frac{T_{step}}{T_{bwd}}$  is constant for a given model and is influenced by factors such as sequence length, batch size, offload communication bandwidth, and CPU processing speed. Longer sequences and larger batch sizes further decrease $\frac{T_{step}}{T_{bwd}}$, increasing the bubble available for T-Offload.

The second requirement is that quantized weight uploads must be completed within the 'bubbles' of the warm-up phase. Utilizing a similar analytical approach, we observe that the forward pass of chunk 2 is delayed by $\lceil \frac{(p-3)}{6} \rceil$ rounds relative to chunk 1. Consequently, we derive:

\begin{equation}
T_{avaliable\_upload}= (p-\lceil \frac{(p-3)}{6} \rceil-1)\frac{T_{fwd}}{2p}\label{eq}
\end{equation}

\begin{equation}
\frac {T_{upload}}{2p} \leq T_{avaliable\_upload}  \Rightarrow  \frac {T_{upload}}{T_{fwd}} \frac {1}{(p-\lceil \frac{(p-3)}{6} \rceil-1)} \leq 1\label{eq}
\end{equation}

Here, $T_{avaliable\_upload}$ represents the idle time during the warm-up phase available for uploading quantized weights, and $T_{upload}$ denotes the total time required to upload quantized weights of all layers. As $P$ increases significantly, this relationship simplifies to $\frac {T_{upload}} {T_{fwd}} \frac{6}{5p} \leq 1$. Similar to the first condition, this requirement becomes easier to meet with increased pipeline stages p and longer sequence lengths. Thus, T-Offload exhibits strong scalability with increasing degrees of pipeline parallelism (PP) and sequence lengths, which is suitable for future development trends.

Additionally, T-Offload scales effectively with increased Data Parallelism (DP). As the level of DP increases, the amount of weights each DP rank needs to update decreases, which in turn reduces the time required for gradient offloading and optimizer computations.

\end{document}